\documentclass[nofootinbib,aps,preprint,superscriptaddress]{revtex4}
\usepackage{amsmath}
\usepackage{citesort}
\usepackage{graphicx}

\newcommand{\beqa}{\begin{eqnarray}}
\newcommand{\eeqa}{\end{eqnarray}}
\newcommand{\beq}{\begin{equation}}
\newcommand{\eeq}{\end{equation}}
\newcommand{\bal}{\begin{align}}
\newcommand{\eal}{\end{align}}
\renewcommand{\Re}{{\cal R}e}
\renewcommand{\Im}{{\cal I}m}

\def\sla{\negthinspace\not\negmedspace}
\def\gsim{\ \rlap{\raise 3pt \hbox{$>$}}{\lower 3pt \hbox{$\sim$}}\ }
\def\lsim{\ \rlap{\raise 3pt \hbox{$<$}}{\lower 3pt \hbox{$\sim$}}\ }

\def\sla{\negthinspace\not\negmedspace}
\newcommand{\Tr}{\operatorname{Tr}}

\newcommand{\fms}[1]{{#1}\!\!\!/}

\newcommand{\n}{\overline{n}}

\newcommand{\w}{\omega}

\newcommand{\SI}{\rm{SCET_I}}
\newcommand{\SII}{\rm{SCET_{II}}}
\newcommand{\T}{\mathcal{T}}

\newcommand{\g}{\gamma}
\newcommand{\mS}{\mathcal{S}}
\newcommand{\C}{\mathcal{C}}
\newcommand{\cP}{\mathcal{P}}

\begin{document}

\baselineskip 3.0ex 

\vspace*{18pt}

\title{Probing electroweak physics using $B\to XM $ decays\\ in the
  endpoint region} 

\def\addKorea{Department of Physics, Korea University, Seoul 136-701,
  Korea} 
\def\addPitt{Department of Physics and Astronomy, University of
  Pittsburgh, PA 15260, USA} 
\def\addcmu{Department of Physics, Carnegie Mellon University,
  Pittsburgh,  PA 15213, USA}
\def\addfmf{Department of Physics, University of Ljubljana, Jadranska
  19, 1000 Ljubljana, Slovenia}
\def\addIJS{J.~Stefan Institute, Jamova 39, P.O. Box 3000, 1001
Ljubljana, Slovenia \vspace{1cm} }

\author{Junegone Chay}\email{chay@korea.ac.kr}\affiliation{\addKorea}
\author{Chul Kim}\email{chk30@pitt.edu}\affiliation{\addPitt} 
\author{Adam K. Leibovich}\email{akl2@pitt.edu}\affiliation{\addPitt} 
\author{Jure Zupan}\email{jure.zupan@ijs.si}
\affiliation{\addfmf}\affiliation{\addIJS}

\begin{abstract} \vspace*{18pt}
\baselineskip 3.0ex 
Using soft-collinear effective theory we describe at leading order in
$1/m_b$ all the semi-inclusive hadronic $B\to XM$ decays near the
endpoint, where an energetic light meson $M$ recoils against 
an inclusive jet $X$. We also include the decays involving
$\eta, \eta'$ mesons that receive additional contributions from gluonic
operators. The predicted branching ratios and CP asymmetries depend on
fewer hadronic parameters than the corresponding two-body $B$
decays. This makes semi-inclusive hadronic $B\to XM$ decays a powerful  
probe of the potential nonperturbative nature of charming penguins as
well as a useful probe of new physics effects in electroweak flavor
changing transitions. A comparison with $B\to KX$ data from BaBar  
points to an enhanced charming penguin, albeit with large experimental
errors.  

\end{abstract}

\maketitle

\section{Introduction} 
Recently BaBar made the first measurement of semi-inclusive $B\to
KX$ branching ratios using fully reconstructed $B$ decays
\cite{Aubert:2006an}
\begin{align}
\label{KBr}
\mathrm{Br}(B^-/\overline{B}^0\to K^-
X)&=(196^{+37+31}_{-34-30})\times 10^{-6}, 
\nonumber \\
\mathrm{Br}(B^-/\overline{B}^0\to \overline{K}^0
X)&=(154^{+55+55}_{-48-41})\times 10^{-6},
\end{align}
where a lower cut on the kaon momentum $p^*(K)>2.34$ GeV in the $B$
rest frame was imposed. This opens up the road  
for experimental explorations in hadronic semi-inclusive $B$
decays, where for almost a decade the only observable probe has been
$\mathrm{Br}(B\to \eta' X)$, first determined by CLEO
\cite{Browder:1998yb}. Averaging over 
the most recent measurements from BaBar \cite{Aubert:2004eq} and CLEO
\cite{Bonvicini:2003aw} gives this branching ratio 
\beq
\mathrm{Br} (B\to \eta'X_s)=(420\pm94)\times 10^{-6}
\eeq
for a lower cut on $\eta'$ energy of $E_{\eta'}>2.218$ GeV.

From the theoretical side semi-inclusive hadronic decays are very
interesting since they are simpler, yet can still probe
many of the questions that have been raised in the context of 
two-body $B$ decays such as the perturbative and nonperturbative
nature of charming penguins \cite{Chay:2006ve} and the search for new
physics signals \cite{Soni:2005jj}. Theoretical simplification
occurs in the endpoint region, where the energy of the
light meson $M$ is relatively close to the maximal energy, so that the
isolated energetic meson $M$ and the inclusive 
collinear hadronic jet $X$ go in opposite
directions. Incidentally, this is also the part of phase space
that is most readily probed experimentally.  

\begin{figure}
\begin{center}
\vspace{0.6cm}
\includegraphics[width=14cm]{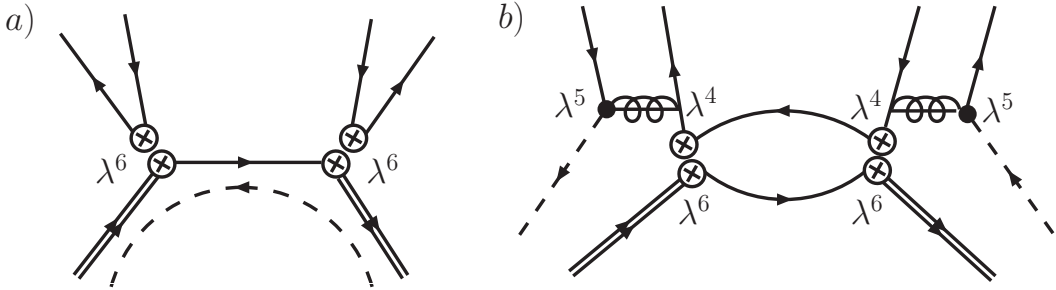}
\end{center}
\caption{\baselineskip 3.0ex  
Time-ordered products of the effective weak operators for the decay
widths: a) the leading-order contributions, b) a subset of subleading
spectator interactions discussed in Section
\ref{sec:powercounting}. The heavy quark fields are  
denoted by double lines, collinear quarks (gluons) by solid lines
(overlaid with wiggly line) and soft quarks by dashed lines. The $n$
hard-collinear quarks connecting the weak vertices and the $\n$
hard-collinear gluons [boosting the spectator in b)] carry $p^2\sim
\Lambda m_b$  and are integrated out. The scaling of vertices is in
$\lambda=\sqrt{\Lambda/m_b}$, cf.~Section \ref{sec:powercounting}.} 
\label{fig:spect}
\end{figure}

First predictions for the semi-inclusive hadronic decays $B\to X M$ in
the endpoint region  were given in
Refs.~\cite{Chay:2006ve,Soni:2005jj} using the soft-collinear
effective theory (SCET)
\cite{Bauer:2000ew,Bauer:2000yr,Bauer:2001ct,Bauer:2001yt} (for
earlier works on semi-inclusive decays using different theoretical
approaches see
\cite{Cheng:2001nj,Atwood:1997de,Browder:1997yq,Eilam:2002wu,Kagan:1997qn,He:1998se,Kim:2002gv,Calmet:1999ix}).   
In the present work we  go beyond Ref.~\cite{Chay:2006ve} in several
ways. First, because of the new experimental data on $B\to K X$
branching ratios in Eq.~\eqref{KBr},  we are able to discuss the
size of charming penguins and include it in perturbative
predictions. Secondly, contrary to Ref.~\cite{Chay:2006ve} in which
only decays  where the spectator quark is part of the inclusive
jet were considered, we extend the discussion to all semi-inclusive
decays, including decays to $\eta,\eta'$. This is simplified by the
fact that contributions where the  spectator quark becomes part of the
exclusive final state meson $M$ are $1/m_b^2 $ suppressed and can be
neglected in our leading order calculations. These 
contributions are schematically shown in Fig.~\ref{fig:spect}b to be
compared with the leading-order contributions in Fig.~\ref{fig:spect}a
(additional gluonic contributions are present for decays into
$\eta,\eta'$).

This means that the nonperturbative parameters $\zeta^{BM},
\zeta_J^{BM}$, connected to the $B\to M$ form factors, do not enter in
the leading order $B\to XM$ predictions, making them simpler 
than the predictions for the corresponding hadronic two-body $B$
decays
\cite{Chay:2003ju,Bauer:2004tj,Williamson:2006hb,Jain:2007dy,Beneke:1999br,Beneke:2005vv,Beneke:2006hg}.  
The presence of 
an inclusive collinear jet in the final state is described by a convolution of
a nonperturbative shape function with a jet function. The latter arises in the 
matching of the full theory onto
$\mathrm{SCET}_{\mathrm{I}}$ at the scale $p_X^2 \sim m_b
\Lambda_{\mathrm{QCD}}$. 
At leading order this
convolution is the same as in $B\to X_s\gamma$ decays, so that many
hadronic uncertainties cancel by taking ratios.

The paper is organized as follows. In Section \ref{sec:powercounting} 
we show the SCET power counting for the different possible decay 
contributions.  This will allow us to include decays where the 
spectator could end up in the meson.  We also discuss in this section the
extra gluonic operators which contribute when the outgoing meson is an 
isosinglet meson.  In Section
\ref{sec:formalism} we briefly review the results of
\cite{Chay:2006ve} and present the hard 
kernels for all semi-inclusive hadronic decays.
In Section \ref{sec:glue} we
discuss the production of $\eta$ and $\eta'$ mesons, where new gluonic
operators are present at leading order in the power counting.
In Section \ref{pheno} we compare the predictions with data and
then conclude in Section \ref{sec:conclusions}.

\section{Power counting}\label{sec:powercounting}
We work at leading order in $1/m_b$ as in Ref.~\cite{Chay:2006ve}. At
this order it is fairly easy to modify the results of
Ref.~\cite{Chay:2006ve} to include the semi-inclusive decays in which
the spectator can go to either the jet or the light meson.  In
particular contributions where the spectator quark is boosted to
become part of the exclusive final state meson $M$ are $1/m_b^2 $
suppressed and can be neglected.  

To show this we first explicitly power count different leading and
subleading graphs in $\SI$, where the expansion parameter is
$\lambda=\sqrt{\Lambda/m_b}$. Order of the graph, $\lambda^\delta$, is
given simply by power-counting 
the different vertices appearing in the graph~\cite{Bauer:2002uv}
\beq
\delta =4 + \sum_k(k-4) V_k,
\eeq
where $V_k$ is the number of vertices that scale as $\lambda^k$ (the
above equation already assumes that there are no purely ultrasoft
diagrams). Now consider the contribution where the spectator ends up
in the jet as shown in Fig.~\ref{fig:spect}a.  For this graph $V_6= 2$
which gives a scaling  
\beq
\delta_{\rm jet} = 4 + (6-4)\times 2 = 8.
\eeq
After $\SI\to\SII$ matching there is an additional suppression of
$\lambda$ for each external collinear line, which results in a final
power counting of $\lambda^{12} = (\Lambda_{\rm QCD}/m_b)^6$.  This is
the leading order term for semi-inclusive hadronic $B$ decays.

\begin{figure}
\begin{center}
\includegraphics[width=6cm]{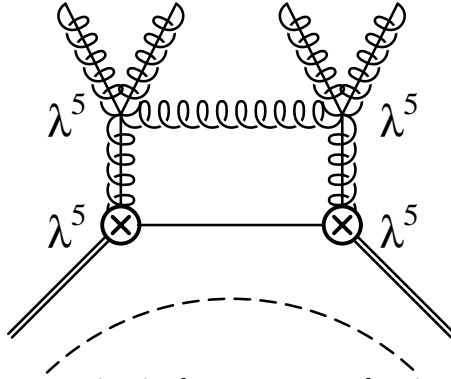}
\end{center}
\vspace{-1.0cm}
\caption{\baselineskip 3.0ex 
Diagram that contributes to isosinglet meson production, which is of the
same order as the leading diagram.}  
\label{fig:isosinglet}
\end{figure}

Next consider the contribution where the spectator ends up in the
meson.  A typical diagram is shown in Fig.~\ref{fig:spect}b with $V_4=
2, V_5= 2$, and $V_6 = 2$, leading to 
\beq
\delta_{\rm spect} = 4 + (4-4)\times 2 + (5-4)\times 2 + (6-4)\times 2= 10.
\eeq
In addition to the usual $\lambda$ suppression
for each external collinear line in the matching $\SI\to\SII$, however,
this diagram is further suppressed due to the $p_\perp$ occurring in
each of the $\lambda^4$ collinear vertices.  Lowering $p_\perp$ to the
$\SII$ scaling gives an extra power of $\lambda$ for both vertices.
This is exactly the same suppression that makes the soft-overlap and
the hard-scattering contributions in heavy-to-light decays of the same
order as discussed in Ref.~\cite{Bauer:2002aj}.  The diagram in
Fig.~\ref{fig:spect}b therefore scales as $\lambda^{16} =
(\Lambda_{\rm QCD}/m_b)^8$ and thus is $1/m_b^2$ suppressed compared
to the leading contribution and can be neglected in the leading order
analysis.  Other possible diagrams in which the spectator ends up in
the final meson give the same suppression and can also be neglected.
Note that this does not mean that all spectator interactions 
are $1/m_b^2$ suppressed. In particular, annihilation contributions
where the spectator quark annihilates with a collinear quark in the
jet arise already at $1/m_b$ order as in $B\to X\gamma$
\cite{Lee:2004ja}.

The decays into isosinglet mesons $\eta, \eta'$ have additional
contributions from gluonic operators such as the one shown in
Fig.~\ref{fig:isosinglet}, for which $V_5 = 4$ and therefore 
\beq
\delta_{\rm iso} = 4 + (5-4)\times 4 = 8,
\eeq
which is the same power suppression as the diagram in
Fig.~\ref{fig:spect}a. After matching onto $\SII$ this diagram then
contributes at leading power, $(\Lambda_{\rm QCD}/m_b)^6$.  Thus, to
analyze isosinglet meson production, we must include new
contributions, complicating the analysis.  Isosinglet meson production 
will be discussed in section \ref{sec:glue}.

\section{The formalism}\label{sec:formalism} 
\begin{figure}[b]
\begin{center}
\vspace{0.6cm}
\includegraphics[width=16cm]{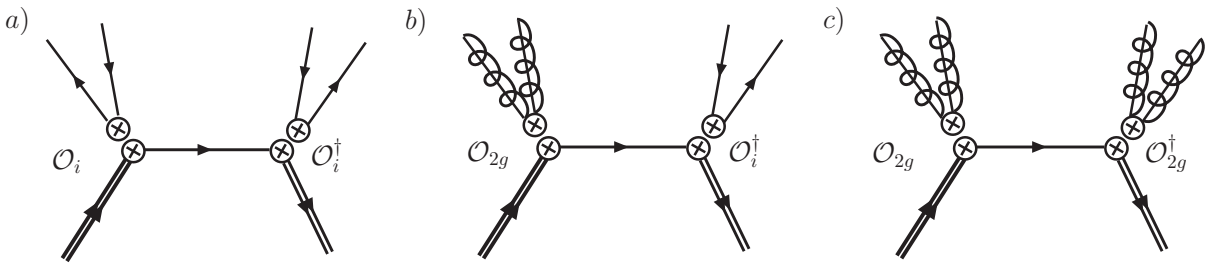}
\end{center}
\caption{\baselineskip 3.0ex  
Time-ordered products of ${\cal O}_i$ effective weak operators giving
the decay widths through use of the optical theorem.  The $n$
collinear quarks connecting the weak vertices carry $p^2\sim \Lambda
m_b$  and are integrated out. The gluonic contributions b) (with an
additional mirror image not shown) and c) contribute only to $B\to
\eta^{(')}X$ decays. }  
\label{Oi_forward} 
\end{figure}

In this section we briefly review the results obtained in
\cite{Chay:2006ve} while extending them to the full set of
semi-inclusive decays. Additional contributions that arise for decays
with $\eta$ or $ \eta'$ in the final state will be included in the
next section. Barring those contributions, the decay rates of 
semi-inclusive $B$ decays are obtained from the forward scattering
amplitude of the time-ordered product of the heavy-to-light currents,
as shown in Fig.~\ref{Oi_forward}. Because of disparate scales in the
problem, a series of matchings on appropriate effective theories is
performed. First, at the scale $\mu\sim m_b$ the standard effective
weak Hamiltonian in full QCD for hadronic $B$ decays
\cite{Buchalla:1995vs} is matched onto the effective Hamiltonian in
$\SI$ by integrating out degrees of freedom of order $m_b$
\cite{Chay:2003ju,Bauer:2004tj,Williamson:2006hb}. In the next step  
$\mathrm{SCET}_{\mathrm{I}}$ is matched onto
$\mathrm{SCET}_{\mathrm{II}}$ by integrating out the degrees of
freedom with  $p^2 \sim m_b \Lambda$ \cite{Chay:2006ve}. As a result,
the jet function is obtained, the discontinuity 
of which contributes to the semi-inclusive hadronic $B$ decay
rates. This jet function is the same as in $B\to X_s \gamma$ and will
cancel once the ratio of decay rates is taken. The predictions for
$B\to X M$ branching ratios normalized to ${\rm Br}(B\to X_s \gamma)$
and for direct CP asymmetries in $B\to X M$ will thus depend only on 
perturbatively calculable hard kernels obtained from matching on
$\mathrm{SCET}_{\mathrm{I}}$ at $\mu\sim m_b$, and on the remaining
nonperturbative parameters - light cone distribution amplitudes (LCDA)
and the parameters describing nonperturbative charming penguins. 

The hard kernels will depend on the Wilson coefficients $\C_{i}^p$ of
the $\mathrm{SCET}_{\mathrm{I}}$ weak Hamiltonian that is at leading
order (LO) in $1/m_b$ given by 
\cite{Chay:2003ju,Bauer:2004tj,Williamson:2006hb} 
\begin{equation} 
H_{I} = \frac{2G_F}{\sqrt{2}} \sum_{p=u,c} \lambda_p^{(q)}
\sum_{i=1}^{6,g} \C_{i}^p \otimes  {\cal O}_i,
\label{hscet}
\end{equation}  
where $\otimes$ denotes the
convolution over collinear momenta fractions, while  $\lambda_p^{(q)}=
V_{pb}V_{pq}^*$ is the CKM factor with $q=s,d$ for $\Delta S=1,0$
transitions. 
The Wilson coefficients ${\cal C}_i^p$ are shown at leading order in
Appendix~\ref{app:tlm}, and were calculated at NLO in $\alpha_s(m_b)$
first in  Refs.~\cite{Beneke:1999br}, and then in 
Ref.~\cite{Chay:2003ju}.  In our notation, the NLO Wilson coefficients
can be found in Appendix A of Ref.~\cite{Chay:2006ve}. 
The sum is over four-quark operators
\begin{alignat}{2}
\label{siop}
{\cal O}_1 &= \big[\overline{u}_n \fms{\overline{n}} P_L 
Y_n^{\dagger} b_v \big] \big[\overline{q}_{\n}  \fms{n} P_L u_{\n}
\big]_u, \qquad\qquad& 
{\cal O}_{2,3} &= \big[\overline{q}_n  \fms{\overline{n}} P_L
Y_n^{\dagger} b_v\big] 
\big[\overline{u}_{\n} \fms{n} P_{L,R} u_{\n}\big]_u,  \\ 
\label{sfquark}
{\cal O}_4 &= \sum_{q'} \big[\overline{q'}_n \fms{\overline{n}} P_L
Y_n^{\dagger} b_v\big] \big[\overline{q}_{\n} \fms{n} P_L
q'_{\n}\big]_u, & {\cal O}_{5,6} &= \sum_{q'} \big[\overline{q}_n
\fms{\overline{n}} P_L Y_n^{\dagger} b_v\big] \big[\overline{q'}_{\n}
\fms{n} P_{L,R} q'_{\n}\big]_u, \nonumber
\end{alignat}
and the gluonic operators are (the trace is over color indices) 
\begin{eqnarray}
\label{QZeroGlue}
{\cal O}_{1g}&=& -\frac{m_b}{4\pi^2} [\overline{q}_{n} Y_n^\dagger
    Y_{\bar     n}\sla \bar n  
n\cdot P ig\sla {\cal B}^{\perp}_{\bar n} P_RY_{\bar n}^\dagger
b_v],\nonumber \\
{\cal O}_{2g}&=&\frac{g^2m_b}{4\pi^2} [\overline{q}_{n}\sla \bar n P_L
    Y_n^\dagger b_v] 
\Tr[{\cal B}^{\perp \mu}_{\bar n}{\cal B}^{\perp \nu}_{\bar
  n}]_ui\epsilon_{\perp \mu\nu}, 
\end{eqnarray}
where the purely gluonic field ${\cal B}^{\perp \mu}_{\bar n}$ is
related to the $(\bar n, \perp)$ component of the gluon field strength
using the usual bracket prescription \cite{Bauer:2004tj} 
\beq
i g {\cal B}^{\perp\mu}_{\bar n}  
= \frac1{n\cdot P}\left[W_{\bar n}^\dagger
[i n\cdot D_{\bar n},\  i D_{\bar  n\perp}^\mu] W_{\bar
  n}\right].  
\eeq
The operators  ${\cal O}_{1g,2g}$ contribute only to the decays with
$\eta, \eta'$ in the final state. We list the operator ${\cal O}_{2g}$
for completeness, in order to have expressions valid to LO in $1/m_b$
but to all order in $\alpha_s(m_b)$.  When we discuss the phenomenology
in section~\ref{pheno}, we work at $O(\alpha_s(m_b))$. At this order  
${\cal O}_{2g}$ has a vanishing matching coefficient
\cite{Williamson:2006hb}  and   
thus does not contribute to the order that we are working.

The summation over $q'$ in Eq.~\eqref{siop} includes $u$, $d$ and $s$
quarks and $P_{L,R} 
= (1\mp \gamma_5)/2$. The notation is the same as the one used in
\cite{Chay:2006ve}. Thus 
$[ \overline{q}_{\n} \fms{n} P_L q_{\n}]_u= [
\overline{q}_{\n}\; \delta\! (u-\frac{n\cdot
  \mathcal{P}^{\dagger}}{2E_M}) \fms{n} P_L q_{\n}]$, 
 while gauge-invariant $n$ and $\bar n$ collinear quark fields
 $q_n=W_n^\dagger \xi_n^{(q)}$ and $q_{\bar{n}}=W_{\bar n}^\dagger
 \xi_{\n}^{(q)}$ already contain the collinear Wilson lines. The
 ultrasoft (usoft)     
Wilson line in the $n$ direction, $Y_n$, arises after the redefinition of
the collinear fields to decouple collinear and usoft degrees of freedom
\cite{Bauer:2001yt}. 

\begin{table}
\begin{tabular}{cccc}\hline\hline
$B^-\to MX $& $\overline B^0\to MX$ & $\overline{B}_s^0\to MX$ &
$T_{M,p}^{(s)}$ \\ \hline  
$K^{(*)-} X_{u\bar u}^0$ &$  K^{(*)-} X_{u\bar d}^+$ & $ K^{(*)-} X_{u
  \bar s}^+$ & $\C_{1}^{p}+ \C_{4}^{p} $  
\\
$\overline{K}^{(*)0} X_{d\bar u}^-$ & $ \overline{K}^{(*)0} X_{d\bar
  d}^0$ &$\overline K^{(*)0} X_{d \bar s}^0$ &  $\C_{4}^{p} $ \\ 
$\phi X_{s\bar u}^-$ &$\phi X_{s\bar d}^0$&  $\phi X_{s\bar s}^0$ &
$\C_{4}^{p}+\C_5+ \C_6$ 
\\ 
$\eta_s X_{s\bar u}^-$ & $  \eta_s X_{s\bar d}^0$& $ \eta_s X_{s\bar s}^0$ &
$\C_{4}^{p}+\C_5- \C_6$ 
\\
$ \omega X_{s\bar u}^-$ & $ \omega X_{s\bar d}^0$& $\w X_{s\bar s}^0$ &
$\big(\C_{2}^{p}+\C_{3} +2\C_5+2\C_6\big)/{\sqrt{2}}$ 
\\
$\eta_q X_{s\bar u}^-$ & $\eta_q X_{s\bar d}^0$& $\eta_q X_{s\bar s}^0$ & 
$\big(\C_{2}^{p}-\C_{3} +2\C_5-2\C_6\big)/{\sqrt{2}}$ 
\\
$\pi^0 X_{s\bar u}^-, \rho^0 X_{s\bar u}^-\,~~~~$ & $\,\pi^0 X_{s\bar
  d}^0, \rho^0 X_{s\bar d}^0\,~~~~$& $\,\pi^0 X_{s\bar s}^0,\rho^0
X_{s\bar s}^0~~~~$ &  
$\big(\C_{2}^{p}\mp \C_3\big)/{\sqrt{2}}$ 
\\
\hline\hline 
\end{tabular}
\caption{\baselineskip 3.0ex 
Hard kernels $\T_{M,p}^{(s)}$ with $p=u,c$ for $\Delta S=1$
semi-inclusive $ B^-/\overline B^0/\overline B^0_s\to X M$ decays.The
NLO Wilson coefficients $\C_i^p$ are given in Appendix A of 
\cite{Chay:2006ve}. For the additional gluonic contributions to decays
with $\eta_{q,s}$ see section \ref{sec:glue}.}
\label{table:TMs}
\end{table}

This decoupling implies that the operators in Eqs.~\eqref{siop} and
\eqref{QZeroGlue} factorize into currents $J_C=(\bar q_{\n} \sla n
\Gamma q'_{\n})$ and $J_H=(\bar q_{n}\sla \bar n P_L Y_n^\dagger b_v)$
which do not exchange soft gluons. The
matrix elements of $\bar n$ currents $\langle M|J_C|0\rangle$ are
expressed in terms of light-cone distribution amplitudes (LCDA), while
the time-ordered product of heavy-to-light currents,  
\begin{equation} 
T(E_M) = \frac{i}{m_B} \int d^4 z ~e^{-ip_M\cdot z} 
\langle B |\mathrm{T} J_H^{\dagger} (z)  J_H (0) | B \rangle, 
\label{tem} 
\end{equation}  
leads to a convolution of shape, $f(l_+)$, and jet functions,
$J_P$, \cite{Chay:2006ve}
\beq \label{Sx} 
\begin{split}
\frac{1}{\pi} \mathrm{Im} \, T(E_M) &=
2 \int^{\overline{\Lambda}}_{-m_b+2E_M} dl_+
~f(l_+) \Bigl[-\frac{1}{\pi}\mathrm{Im}J_P \Bigl(m_b-2E_M +
l_+ + i\epsilon\Bigr) \Bigr] \\
&\equiv \frac{2}{m_b} \mS(E_M,\mu_0),
\end{split}
\eeq
where $l_+=n \cdot l$ is the soft momentum conjugate to the $\bar
n\cdot z=z_-$ spatial component. 
Using the optical theorem this is then related to the $B\to X M$ decay
rate giving  
\begin{equation}  
\frac{d\Gamma}{dE_M}(B\to X M) = \frac{G_F^2}{8\pi} m_b^2 x_M^3
\mS(x_M, \mu_0) \big|h_M^{(q)}\big|^2+\cdots,  
\label{decay} 
\end{equation} 
where $x_M=2 E_M/m_b\simeq 1$. The ellipses represent nonperturbative
charming penguin contributions given explicitly below, while
$h_{M}^{(q)}$ is the convolution of the hard kernel and the LCDA 
\beq
h_{M}^{(q)}=f_M \int_0^1 du\, \phi_M(u) \big[\lambda_u^{(q)}
{T}_{M,u}^{(q)}(u)+\lambda_c^{(q)} {T}_{M,c}^{(q)}(u)\big]. 
\eeq 
Here $\phi_M(u)$ is the light meson LCDA, $f_M$ the decay constant,
$\lambda_p^{(q)}=V_{pb}V_{pq}^*$ the CKM elements, while the
perturbatively calculable hard kernels ${T}_{M,p}^{(q)}$ are given in
Tables \ref{table:TMs} and \ref{table:TMd} for $\Delta S=1,0$
$(q=s,d)$, respectively.  

\begin{table}
\begin{tabular}{cccc}\hline\hline
$B^-\to MX$& $\overline B^0\to MX$ & $\overline{B}_s^0\to MX$ &
$T_{M,p}^{(d)}$ \\ \hline  
$\pi^-X_{u\bar u}^0,\rho^- X_{u\bar u}^0~~~$ &$\pi^-X_{u \bar
  d}^+,\rho^- X_{u \bar d}^+~~~$& $\pi^{-} X_{u\bar s}^+,
\rho^{-}X_{u\bar s}^+~~~$  &  $ \C_{1}^{p}+ \C_{4}^{p} $   
\\ 
$\pi^0X_{d\bar u}^-,\rho^0 X_{d\bar u}^-$ & $\pi^0 X_{d\bar
  d}^0,\rho^0 X_{d\bar d}^0$  &  $\pi^0 X_{d \bar s}^0,\rho^0 X_{d
  \bar s}^0$& 
$\big(\C_{2}^{p}-\C_{4}^{p}\mp \C_{3}\big)/\sqrt{2} $  
\\ 
$K^{(*)0}X_{s\bar u}^-$ & $ K^{(*)0}X_{s\bar d}^0$ & $
K^{(*)0}X_{s\bar s}^0$ &$\C_{4}^{p} $  \\  
$\omega X_{d\bar u}^-$ & $ \omega X_{d\bar d}^0$&  $\w X_{d\bar s}^0$ &
$\big(\C_{2}^{p}+\C_{4}^{p}+\C_{3} +2\C_5+2\C_6\big)/{\sqrt{2}}$ 
\\
$\eta_q X_{d\bar u}^-$ & $ \eta_q X_{d\bar d}^0$&  $\eta_q X_{d\bar
  s}^0$ & 
$\big(\C_{2}^{p}+\C_{4}^{p}-\C_{3} +2\C_5-2\C_6\big)/{\sqrt{2}}$  \\
 $\phi X_{d\bar u}^-$ & $\phi X_{d\bar d}^0 $& $\phi X_{d\bar s}^0$ &
 $\C_5+ \C_6$ \\  
 $\eta_s X_{d\bar u}^-$ & $\eta_s X_{d\bar d}^0 $& $\eta_s X_{d\bar
   s}^0$ &  $\C_5- \C_6$ \\ 
\hline\hline 
\end{tabular}
\caption{\baselineskip 3.0ex 
Hard kernels $\T_{M,p}^{(d)}$ with $p=u,c$ for $\Delta S=0$
semi-inclusive $B^-/\overline B^0/\overline B^0_s\to X M$ decays. The
summation over $p=u,c$ is implied. The NLO Wilson 
coefficients $\C_i^p$ are given in Appendix A of \cite{Chay:2006ve}.
For the additional gluonic contributions to decays
with $\eta_{q,s}$ see section \ref{sec:glue}.}
\label{table:TMd}
\end{table}

The nonperturbative function $\mS$ denoting the convolution of shape
and jet functions is exactly the same as the one appearing in
the prediction for the $B\to X_s\gamma$ rate in the endpoint region at LO
in $1/m_b$. In the ratio 
with $\Gamma(B\to X_s\gamma)$ it thus cancels out, giving 
\beq
\begin{split}\label{ratioG}
&\frac{d\Gamma(\bar B\to M X)/dE_M}{d\Gamma(\bar B\to X_s
  \gamma)/dE_\gamma}=\frac{2\pi^3}{\alpha m_b^2}\; 
\frac{\Big( \big|h_M^{(q)}\big|^2+2 \Re\big[\lambda_c^{(q)}
 c_{cc}\, {p}_{cc}^M \big(h_M^{(q)}\big)^* \big]+\big|
  \lambda_c^{(q)}c_{cc}\big|^2
 {\cP}_{cc}^M\Big)}{|\lambda_t^{(s)}C_\gamma(c_9^{\rm eff}+1/2
  c_{12}^{\rm eff})|^2}, 
\end{split}
\eeq
where one sets $E_\gamma=E_M$. The SCET Wilson coefficients are
$c_9^{\rm eff}=1$, $c_{12}^{\rm eff}=0$ at LO with the NLO calculated
in \cite{Bauer:2000yr}, while $C_\gamma$ is given  
e.g.~in Eq.~(13) of  \cite{Lee:2004ja}.

\begin{figure}
\begin{center}
\vspace{0.6cm}
\includegraphics[width=15cm]{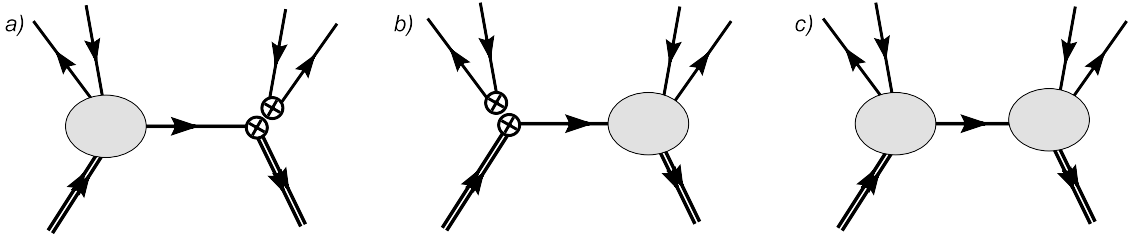}
\end{center}
\caption{\baselineskip 3.0ex  
Charming penguin contributions: the contributions from a) and b)
are proportional to $p_{cc}^M$ and $p_{cc}^{M*}$ respectively, while
the contribution from  c) gives ${\cal P}_{cc}^M$. The blobs represent
nonperturbative charming penguins}  
\label{fig:charm_peng}
\end{figure}

The coefficients $p_{cc}^M$ and $\cP_{cc}^M$  parametrize possible
nonperturbative charming penguin contributions.\footnote{In
  \cite{Soni:2005jj} these were $p_{cc}^M=f_M\bar {f}_{cc}$ and
  $\cP_{cc}^M=f_M^2\overline {\cal F}_{cc}$.} They depend on the
valence quark structure of both $M$ and the jet $X$, but we suppress this
dependence in the notation. They are zero if the
charming penguin contributions are purely perturbative. However, the
uncalculated higher-order perturbative pieces can mimick their effect,
making them differ slightly from zero.
The complex parameter $p_{cc}^M$ describes the interference of the
nonperturbative charming penguin with the perturbative hard kernels,
shown in Fig.~\ref{fig:charm_peng}a. 
The positive real parameter $\cP_{cc}^M$ in Eq.~\eqref{ratioG} on the
other hand describes the square of the nonperturbative charming
penguin contributions shown in Fig.~\ref{fig:charm_peng}c. 
If hard kernels dominate the amplitudes, the term with $p_{cc}^M$  in
Eq.~\eqref{ratioG} is subleading, while the $\cP_{cc}^M$ term is even
more suppressed and can be neglected as was done in
\cite{Chay:2006ve}. It should be kept, however, if nonperturbative
charming penguins are sizable. Since the present data are inconclusive
we keep both terms in Eq.~\eqref{ratioG}. When estimating the size of
the non-perturbative contributions, as a rule of thumb we will take
$(p_{cc}^{M})^2 \sim \cP_{cc}^M$. Further information  
on the structure of  $p_{cc}^{M}, \cP_{cc}^M$ can be obtained in the
$m_c\to \infty$ limit \cite{Chay:2006ve}. 
Finally, the coefficient $c_{cc}$
multiplying the nonperturbative charming penguin parameters in
Eq.~\eqref{ratioG} is equal to the coefficient of $\C_4^c$ in Tables
\ref{table:TMs} and \ref{table:TMd} (i.e., it is  $c_{cc}=1$
for $B^-\to \pi^- X_{u \bar u}^0$ and $c_{cc}=-1/\sqrt{2}$ for $B^-\to
\pi^0 X_{d \bar u}^-$ so that $p_{cc}^M$ in both cases equals
$p_{cc}^{\pi}$).    

In the phenomenological analysis of our results in Section~\ref{pheno} we
give numerical estimates for direct CP asymmetries  
\beq\label{ACP}
A_{CP}(\overline{B}\to XM)=\frac{{d\Gamma(\overline{B}\to XM)}/{dE_M}
  - d\Gamma(B\to 
  XM)/dE_M}{d\Gamma(\overline{B}\to XM)/dE_M + d\Gamma(B\to XM)/dE_M} ,
\eeq
and CP averaged branching ratios. In terms of hard kernels and
nonperturbative charming penguin parameters the direct CP asymmetry is 
\beq\label{ACP2}
A_{CP}(\overline{B}\to XM)=\frac{-2
  \Im\big(\lambda_u^{(q)}\lambda_c^{(q)*}\big)}{|A(\overline{B}\to
  XM)|^2} \Im\big[T(P+c_{cc}p_{cc}^M)^*\big], 
\eeq
and the CP averaged decay width normalized to $B\to X_s\gamma$ is
\beq
\begin{split}\label{ratioGCP}
&\frac{d\Gamma_{CP}(\bar B\to M X)/dE_M}{d\Gamma(\bar B\to X_s
  \gamma)/dE_\gamma}=\frac{2\pi^3}{\alpha m_b^2}\; 
\frac{|A(\overline{B}\to XM)|^2}{|\lambda_t^{(s)}C_\gamma(c_9^{\rm
    eff}+1/2 c_{12}^{\rm eff})|^2}, 
\end{split}
\eeq
where
\beq
\begin{split}
|A(\overline{B}\to XM)|^2=&|\lambda_u^{(q)}|^2|T|^2+2
\Re\big(\lambda_u^{(q)}
\lambda_c^{(q)*}\big)\Re\big[T(P+c_{cc}p_{cc}^M)^*\big]\\   
&+|\lambda_c^{(q)}|^2\big[|P|^2+2\Re\big(c_{cc}p_{cc}^M
P^*\big)+c_{cc}^2 \cP_{cc}^M\big]. 
\end{split}
\eeq
Above a shorthand notation for the perturbative ``tree" and ``penguin"
contributions 
\beq
T=\int_0^1 du f_M \phi_M(u) T_{M,u}^{(q)}(u), \qquad P=\int_0^1 du f_M
\phi_M(u) T_{M,c}^{(q)}(u), 
\eeq
has been used, dropping in the notation the dependence on
$\overline{B}\to XM$.

\section{The decays involving $\eta, \eta'$}\label{sec:glue} 
\begin{figure}[t]
\begin{center}
\vspace{0.6cm}
\includegraphics[width=14cm]{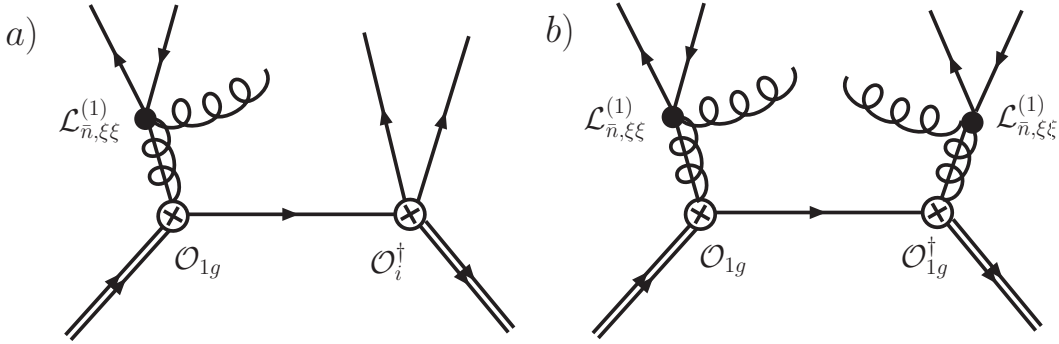}
\end{center}
\caption{\baselineskip 3.0ex  
The contributions of ${\cal O}_{1g}$ operator to $B \to M X_q$
decays. The mirror image of Diagram a) is not shown as well as not the 
diagrams with ${\cal L}_{\bar n,\xi\xi}^{(1)}$ replaced by ${\cal
  L}_{\bar n,cg}^{(1)}$, cf.~Fig.~\ref{glueto}. }   
\label{semigluonic} 
\end{figure}

In order to describe $B\to \eta^{(')}X$ decays several modifications
of the results in the previous section are needed: (i) $\eta-\eta'$
mixing needs to be taken into account and (ii) there are additional
contributions from gluonic operators ${\cal O}_{1g,2g}$ as shown in
Figs.~\ref{Oi_forward}, \ref{semigluonic} [operators ${\cal
  O}_{1g,2g}$ are defined in Eq.~\eqref{QZeroGlue}].  
To describe matrix elements involving gluonic operators we introduce
the gluonic LCDA  
\cite{Kroll:2002nt,Blechman:2004vc}
\beq\label{gluonicwave}
i\epsilon_{\perp \mu \nu} \langle P(p) | \Tr[{\cal B}_{\bar n}^{\perp
    \mu} {\cal B}_{\bar n}^{\perp
    \nu}]_u|0\rangle=\frac{i}{4}\sqrt{C_F} f_P^1 \bar \Phi_P^g(u), 
\eeq 
where  the isosinglet decay constant is the same one that appears in
the matrix elements of quark bilinears   
\beq\label{quarkwave}
\langle P(p) | [\overline{q}_{\bar n} \sla n \gamma_5 T_{1,8}q_{\bar
  n}]_u|0\rangle=-2i E f_P^{1,8}  \phi_P^{1,8}(u), 
\eeq 
with $T_8=\lambda_8/\sqrt{2}$, $T_1=1/\sqrt{3}$ diagonal $3\times 3$
matrices in $u,d,s$ flavor space. The flavor singlet LCDA
$\phi_P^{1}(u)$ mixes with the gluonic LCDA 
$\bar \Phi_P^g(u)$ under RG running \cite{Chase:1980hj}, while
$\phi_P^{8}(u)$ does not. For future reference we also quote 
explicitly $f_{\eta_q}^1=\sqrt{2/3}f_{\eta_q}$,
$f_{\eta_s}^1=f_{\eta_s}/\sqrt{3}$, while $f_P^1=0$ for other
pseudoscalars that do not have flavor singlet component. Here
$f_{\eta_q}$ and $f_{\eta_s}$ are the decay constants corresponding 
to $\bar q q=(\bar u u+\bar d d)/\sqrt{2}$ and $\bar s s$ axial currents 
respectively (and are equal in the SU(3) limit, cf.~also \eqref{feta}
below). 

The parametrizations of matrix elements \eqref{gluonicwave},
\eqref{quarkwave} do not involve any assumptions; however they are too
general for 
the limited amount of data available at present. To reduce the number
of unknowns we use  the FKS mixing scheme \cite{Feldmann:1998vh} to
describe $\eta-\eta'$ mixing in 
which the mass eigenstates $\eta$, $\eta'$ are related to the flavor
basis through 
\beq
\eta=\eta_q \cos \varphi -\eta_s \sin \varphi, \quad
\eta'=\eta_q \sin \varphi +\eta_s \cos \varphi, 
\label{mixing}
\eeq
with $\varphi=(39.3\pm1.0)^\circ$ and
$\eta_q=(\eta_u+\eta_d)/\sqrt{2}$. The working assumptions of the FKS
scheme are that LCDA do not depend on the meson so that 
\beq
 \phi_P^{1,8}(u) =\phi^{1,8}(u), \qquad \bar \Phi_P^g(u)=\bar \Phi^g(u),
 \eeq
and that OZI suppression is effective. This last requirement is most
transparent in the $\eta_q,\eta_s, g$ basis instead of the $1,8,g$
basis used above. In it we have 
\beq
\begin{split}\label{feta}
\langle \eta_q(p) |&  \tfrac{1}{\sqrt{2}}([\bar u_{\bar n} \sla n
\gamma_5 u_{\bar n}]_u+[\bar d_{\bar n} \sla n \gamma_5 d_{\bar n}]_u)
|0\rangle=-2i E f_{\eta_q}  \phi_{\eta_q}(u),\\ 
\langle \eta_s(p) |& [\bar s_{\bar n} \sla n \gamma_5 s_{\bar n}]_u
|0\rangle=-2i E f_{\eta_s}  \phi_{\eta_s}(u), 
\end{split}
\eeq
with $\phi_{\eta_s}(u)=[2\phi^{8}(u)+\phi^{1}(u)]/3$ and
$\phi_{\eta_q}(u)=[2\phi^{1}(u)+\phi^{8}(u)]/3$. The OZI suppressed
matrix elements on the other hand are
\beq
\begin{split}\label{OZI}
\langle \eta_s(p) |& \tfrac{1}{\sqrt{2}}([\bar u_{\bar n} \sla n
\gamma_5 u_{\bar n}]_u+[\bar d_{\bar n} \sla n \gamma_5 d_{\bar n}]_u)
|0\rangle=-2i E f_{\eta_s}  \phi_{\rm opp}(u),\\ 
\langle \eta_q(p) |& [\bar  s_{\bar n} \sla n \gamma_5 s_{\bar n}]_u
|0\rangle=-2i E f_{\eta_q}  \phi_{\rm opp}(u), 
\end{split}
\eeq
where $\phi_{\rm opp}(u)=\sqrt{2}[\phi^{1}(u)-\phi^{8}(u)]/3$ and is
negligible as long as $\phi^{1}(u)\simeq \phi^{8}(u)$. This relation
is exact for asymptotic forms of LCDA, while it can only be
approximate for physical values of $\mu$ since $\phi^{1}(u)$ and
$\phi^{8}(u)$ have different RG runnings, spoiling the relation for
smaller values of $\mu$. Phenomenologically, however, for $\mu$ above
$1 $ GeV the relation is well obeyed at a percent level
\cite{Kroll:2002nt}.

We are now ready to write down the results for contributions to $B\to
\eta (\eta')X$ decays corresponding to Figs.~\ref{Oi_forward} and
\ref{fig:charm_peng}. These are described by Eq.~\eqref{ratioG} but
with ${h}_{M}^{(q)}$ and charming penguin parameters as given below.  
Utilizing the FKS scheme with Eq.~\eqref{feta} and setting the OZI
suppressed matrix elements Eq.~\eqref{OZI} to zero, the $h_M^{(q)}$
functions in Eq.~\eqref{ratioG} are  
\begin{align}\label{heta}
{h}_{\eta}^{(q)}=& \cos \varphi f_{\eta_q} \phi_{\eta_q}\otimes
\lambda_p^{(q)}{T}_{\eta_q,p}^{(q)}-\sin \varphi f_{\eta_s}
\phi_{\eta_s}\otimes \lambda_p^{(q)} {T}_{\eta_s,p}^{(q)}
+h_{\eta}^{g},\\ 
{h}_{\eta'}^{(q)}=& \sin \varphi f_{\eta_q} \phi_{\eta_q}\otimes
\lambda_p^{(q)}{T}_{\eta_q,p}^{(q)}+\cos \varphi f_{\eta_s}
\phi_{\eta_s}\otimes \lambda_p^{(q)}{T}_{\eta_s,p}^{(q)}+h_{\eta'}^g, 
\end{align} 
where $\otimes$ denotes a convolution, while
${T}_{\eta_{q,s},p}^{(q)}$ are listed in Tables \ref{table:TMs},
\ref{table:TMd} and the sum over $p=u,c$ is understood. 
The gluonic contributions $h_{M}^g$ coming from the ${\cal
  O}_{2g}$ operator insertions as shown in Figs.~\ref{Oi_forward}b,
\ref{Oi_forward}c, are zero to NLO in $\alpha_s(m_b)$, i.e.~to the
order we are working. Explicitly, they are 
\beq
h_{\eta_{s,q}}^g
=-\lambda_t^{(s)} \frac{\sqrt{C_F}}{2}
f_{\eta_{s,q}}^1 \bar \Phi_{\eta_{s,q}}^g(u)\otimes \C_{2g}(u), 
\eeq
for $\Delta S=1$ decays $B_{q'}\to \eta_{s,q} X_{s\bar q'}$, while for
$\Delta S=0 $ decays $B_{q'}\to \eta_{s,q} X_{d\bar q'}$ we need to replace
$\lambda_t^{(s)}\to \lambda_t^{(d)}$. The
expressions for $\eta,\eta'$ final states are easily obtained using
\eqref{mixing}. As already stated, $\C_{2g}(u)=0$ at NLO in
$\alpha_s(m_b)$.

For the charming penguin parameters in Eq.~\eqref{ratioG}, we make the
following replacements for $\Delta S=1$ transitions  
\beq 
\begin{split}
 c_{cc} p_{cc}^{\eta}\to-\sin\varphi\,
p_{cc}^{\eta_s},\quad &c_{cc}^2 \cP_{cc}^{\eta}\to \sin^2 \varphi
 \cP_{cc}^{\eta_s}, \\ 
c_{cc}p_{cc}^{\eta'}\to \cos\varphi\,
p_{cc}^{\eta_s},\quad &c_{cc}^2 \cP_{cc}^{\eta'}\to
\cos^2 \varphi \cP_{cc}^{\eta_s},  
\end{split}\label{pccDeltaS=1}
\eeq
and for $\Delta S=0$ transitions
\beq
\begin{split}
c_{cc} p_{cc}^{\eta}\to \frac{\cos\varphi}{\sqrt{2}}
 p_{cc}^{\eta_q},\quad 
c_{cc}^2\cP_{cc}^{\eta}\to \frac{\cos^2 \varphi}{2}
\cP_{cc}^{\eta_q},\\ 
c_{cc} p_{cc}^{\eta'}\to \frac{\sin\varphi}{\sqrt{2}}
 p_{cc}^{\eta_q} ,\quad  c_{cc}^2
\cP_{cc}^{\eta'}\to \frac{\sin^2 \varphi}{2}
\cP_{cc}^{\eta_q}.\label{peta'c} 
\end{split}
\eeq

\begin{figure}
\begin{center}
\vspace{0.6cm}
\includegraphics[width=11cm]{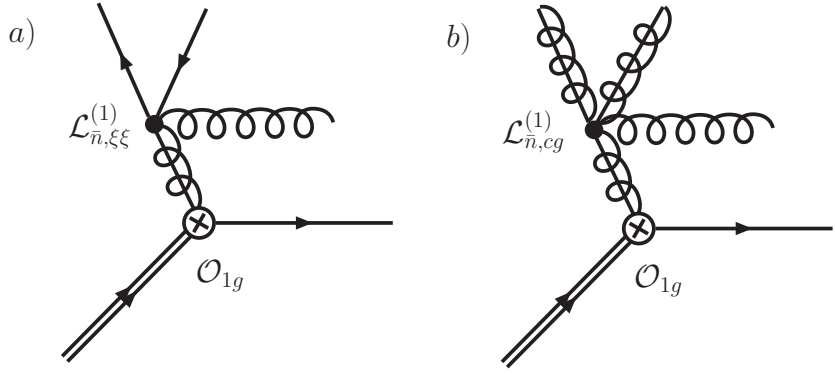}
\end{center}
\caption{\baselineskip 3.0ex  
Time-ordered products between $\mathcal{O}_{1g}$ and the subleading
interaction terms in SCET Lagrangian. The soft gluon (curly line) is
absorbed in a definition of new $B$ meson shape function $f_g$. The
intermediate hard-collinear gluon has offshellness $\sim m_b \Lambda$
and is integrated out. At LO in $\alpha_s(\sqrt{\Lambda m_b})$ Diagram
b) does not contribute because of the antisymmetry of gluonic LCDA. }  
\label{glueto}
\end{figure}

We next move to the contributions from the ${\cal O}_{1g}$ operator,
shown in Fig.~\ref{semigluonic}. These contributions lead to a
modified factorization between $n$ and $\bar n$ degrees of freedom
because of the additional soft gluon that is emitted at a light-like
separation from the weak vertex 
in the $\bar n$ direction.
Relegating the details to Appendix \ref{app:Gderiv}, we quote here
only the main results starting  
with the $T$ products in Fig.~\ref{glueto}
\beq\label{O1gtprod-maintext}
{\cal G}_{\xi\xi (cg)}=\langle M X|\;i
\negthickspace\int\negthickspace d^4 x \; T\big\{{\cal O}_{1g}(0), 
{\cal L}_{\xi\xi(cg)}^{(1)} (x)\big\}|B\rangle. 
\eeq
Then ${\cal G}_{\xi\xi}+{\cal G}_{cg}$ describes the contribution of
the operator ${\cal O}_g$ to the decay into a color-singlet state.

In the $\SI\to \SII$ matching the intermediate hard-collinear gluon
carrying $p^2\sim \Lambda m_b$ is integrated out 
leading respectively to the jet functions $J_1(u, k_-)$ and
$J_g(u,k_-)$ for the diagrams in Fig.~\ref{glueto}a and \ref{glueto}b.  
Following the usual redefinition of fields the $\bar n$ collinear
quark (gluon) lines decouple from the soft fields \cite{Bauer:2001yt}
and lead to quark (gluon) LCDA after taking the matrix element. The soft
Wilson lines $Y_{\bar n}$ arising from the field redefinition and the
soft gluon field emitted at $x$ are taken to be part of the
heavy-to-light current 
\beq\label{JH-maintext}
\tilde J_H(0,x_+)= \overline{q}_n    Y_n^\dagger Y_{\bar n}(0) \sla
\bar n g  \sla {\cal A}_{\rm us}^{\perp}(x_+)P_R Y_{\bar n}^\dagger
b_v(0),  
\eeq
where we define
\beq
g   {\cal A}_{\rm us}^{\perp\mu}=\big[Y_{\bar n}^\dagger i
D_{\rm us}^{\perp\mu} Y_{\bar n}\big]. \label{A:def}
\eeq
Note that the heavy-to-light current $\tilde{J}_H$ depends on $x_+$,
since the soft gluon gets emitted away from the  
 weak vertex.  The ${\cal O}_{1g}$ contribution to the $B\to MX$
 matrix element then  takes a form of a convolution over both the soft
 momentum  $k_-$ (conjugate to the position $x_+$) and the hard
 momentum fraction $u p_M$, 
\beq\label{tildeJ-maintext}
{\cal G}_{\xi\xi}+{\cal G}_{cg}=-i\int du\int \frac{d k_-
  dx_+}{4\pi} e^{-i k_- x_+/2} F^M(k_-,u) \langle X|\tilde
J_H(0,x_+)|B\rangle. 
\eeq
The hard-collinear kernel multiplied by the LCDA is explicitly
\beq
F^M(k_-,u)=\frac{\alpha_s m_b}{4\pi}\left\{ f_M \phi_M(u)
\left[\frac{J_1(u, k_-)}{u} -\frac{J_1(\bar u, -k_-)^*}{\bar u}
  \right]+\sqrt{C_F} f_M^1  \bar \Phi_M^g(u) J_g(u, k_-)\right\},
\label{F^M} 
 \eeq
and is in general both a function of the $\bar n$ momentum fraction $u$ and
the soft momentum $k_-$. At tree level, however, it is a simple
product of functions that depend only on $u$ and only on $k_-$,
\beq
F^M(k_-,u)\big|_{\rm tree}=\left.\frac{\alpha_s m_b}{4\pi} f_M \phi_M(u)
\big(\frac{1}{u}+\frac{1}{\bar u}\big) J_1(k_-) \right|_{\rm tree}, 
\label{simple_prod} 
\eeq
since $J_1(u,k_-)|_{\rm tree}=J_1(k_-)|_{\rm tree}=1/(N k_-)$ and
$J_g(u,k_-)|_{\rm tree}=1/k_-$ are independent of $u$. Furthermore,  
since $\bar \Phi_P^g(u)$ is antisymmetric, $\bar \Phi_P^g(u)=-\bar
\Phi_P^g(\bar u)$, the contribution  
from $J_g(u,k_-)$ vanishes at this order.

Therefore at least at leading order in $\alpha_s(\sqrt{\Lambda m_b})$ 
the additional gluonic contributions can be cast in the same form as
the expressions for the decay widths for nonisosinglet
final states, Eq.~\eqref{decay}, by moving the $u$-dependent part into
the definition of 
hard kernels, while including the dependence on $k_-$ in the
definition of the modified heavy-to-light current
\beq
\tilde {\cal J}(0)=\int \frac{d k_- dx_+}{4\pi} e^{-i k_- x_+/2}
 \tilde J_H(0,x_+)\left. J_1(k_-)\right|_{\rm
   tree}. \label{tildeJ-integrated} 
\eeq
Because of this simplification we will show in this section only the
result for $B\to XM$ decay width at leading order in
$\alpha_s(\sqrt{\Lambda m_b})$, while the result valid to all orders
in $\alpha_s(\sqrt{\Lambda m_b})$ is given in Appendix \ref{app:Gderiv}.

As in Section \ref{sec:formalism} we relate the $B\to XM$ decay width
to the time-ordered product of heavy 
currents using the optical theorem. We denote the $T$-product coming from a
single ${\cal O}_{1g}$ insertion, shown in Fig.~\ref{semigluonic}a, as
\beq\label{Tg}
\tilde T_g(E_M)=\frac{i}{m_b}\int d^4 z \langle  B| T
J_H^\dagger(z) \tilde {\cal J}(0) | B\rangle, 
\eeq
where $J_H(z)=e^{i(\tilde p-m_b v)\cdot z}(\bar q_n
\sla \bar n P_L Y_n^\dagger b_v)(z)$. For the contribution coming from
two insertions of ${\cal O}_{1g}$, shown in  
Fig.~\ref{semigluonic}b, we similarly define  
\beq\label{Tgg-text}
\tilde T_{gg}(E_M)=\frac{i}{m_b}\int d^4 z \langle  B| T \tilde {\cal
  J}^\dagger(z) \tilde {\cal J}(0)| B\rangle. 
\eeq

In evaluating the time ordered product $\tilde T_g(E_M)$ we
use the fact that $n$ collinear quark fields do not exchange any soft
gluons at LO in $1/m_b$ with the other fields in the
$T$-product. Using the standard definition of the $n$-collinear jet
function, 
\beq\label{jet-func}
\langle 0|T q_n(z) \bar q_n(0)|0\rangle =i \frac{\sla n}{2}
\delta(z_+)\delta^2(z_\perp)\int \frac{d \kappa_+}{2\pi}e^{-i \kappa_+
  z_-/2} J_P(\kappa_++i\epsilon), 
\eeq
and a shape function that, unlike $f(l_+)$ in Eq.~\eqref{Sx}, depends on two
soft momenta because of the additional nonlocal structure present in
$\tilde {\cal J}$ due to ${\cal A}_{\rm us}$,
\beq
\begin{split}\label{f_g2}
\int dl_+ e^{i l_+z_-/2}&\int dr_- e^{i r_- x_+/2} f_g(l_+,r_-)=\\
&\langle  B_v| \bar b_v Y_n(z_-)Y_n^\dagger Y_{\bar n}(0)\sla \n g
 \sla {\cal A}_{\rm us}^\perp (x_+)P_R Y_{\bar n}^\dagger b_v(0)
 |B_v\rangle, 
 \end{split}
 \eeq
 the discontinuity of $\tilde T_g(E_M)$ with respect to the
 intermediate states is given by
 \beq
 \begin{split}\label{tildefg}
 {\rm Disc.} \  \tilde T_g(E_M) &= 2 \int dl_+ dr_- \Im
 \left[-\frac{1}{\pi}J_P(l_++m_b-2E_M +i \epsilon) \right] f_g(l_+,r_-)
 J_1(r_-)|_{\rm tree}\\ 
 &\equiv\frac{2}{m_b} {\cal S}_g(E_M,\mu_0).
 \end{split}
 \eeq
Note that the new shape function $f_g(l_+,r_-)$ can be in general complex, hence
${\cal{S}}_g$ can also be complex. However for decays rates, the complex conjugate should be added and the decay rates becomes real.

Using the optical theorem the contribution to the decay width
from one insertion of ${\cal O}_{1g}$  at LO in
$\alpha_s(\sqrt{\Lambda m_b})$, with the contribution of the mirror image of Fig.~\ref{semigluonic}a, is therefore 
\begin{eqnarray}
\Big(\frac{d\Gamma}{dE_M}\Big)_g\Big|_{\rm tree}&=&\frac{G_F^2}{4\pi}
m_b^2 x_M^2 f_M^2  \\
&\times& 2\Re\left\{\left(\phi_M\otimes\lambda_p^{(q)}
T_{M,p}^{(q)}\right)^* \left[ \phi_M \otimes \lambda_t^{(q)}
  \C_{1g} \frac{\alpha_s}{4\pi} \left(\frac{1}{u}+\frac{1}{\bar
      u}\right)\right] {\cal S}_g(E_M,\mu_0) \right\},  \nonumber\label{one-decay}    
\end{eqnarray}
where ${\cal{C}}_{1g}$ is the Wilson coefficient of ${\cal{O}}_{1g}$,
which is 1 at leading order. 
This expression is similar to Eq.~\eqref{decay}. The hard kernels
(convoluted with LCDA) in the curly brackets do not  
depend on soft momenta $k_-$ and similarly the nonperturbative
``shape" function ${\cal S}_g$ does not 
depend on the large momenta fractions $u$. This factorization of the
$u$ and $k_-$ dependence is a consequence 
of a special form of $F^M$ at leading order in $\alpha_s(\sqrt{\Lambda
  m_b})$, Eq.~\eqref{simple_prod}, and may not be  
present at higher orders, cf.~Eq.~\eqref{F^M}.

The contribution $\tilde T_{gg}$, Eq.~\eqref{Tgg-text}, coming from
two insertions of ${\cal O}_{1g}$ leads to a shape function that  
depends on three soft momenta because of two soft gluon insertions
\beq
\begin{split}
\int dl_+ e^{i l_+z_-/2}&\int dr_- e^{i r_- x_+/2} \int ds_- e^{i s_-
  y_+/2}f_{gg}(l_+,r_-,s_-)=\\ 
&\langle  B_v| \bar b_v Y_{\n}(z_-) g \sla {\cal A}_{\rm us}^\perp
(y_+) Y_{\bar n}^\dagger Y_{n}(z_-) Y_n^\dagger Y_{\bar n}(0)\sla \n g
\sla {\cal A}_{\rm us}^\perp (x_+)P_R Y_{\bar n}^\dagger b_v(0)
|B_v\rangle. \label{f-three}
\end{split}
\eeq
This gives a new nonperturbative ``shape'' function by taking the
discontinuity   
\beq
\begin{split}\label{fgg}
{\rm Disc.}\  \tilde T_{gg}(E_M) &= 2 \int dl_+ dr_-ds_-\Im
\Bigl[-\frac{1}{\pi}J_P(l_++m_b-2E_M +i \epsilon) \Bigr] \\ 
&\times J_1(r_-) J_1(s_-)^* f_{gg}(l_+,r_-,s_-)
\equiv\frac{2}{m_b}  {\cal S}_{gg}(E_M,\mu_0), 
\end{split}
\eeq
which then enters the prediction for the contribution of double ${\cal
  O}_{1g}$ insertion to the decay width 
\beq
\Big(\frac{d\Gamma}{dE_M}\Big)_{gg}\Big|_{\rm tree}=\frac{G_F^2}{2\pi}
m_b^2 x_M f_M^2 {\cal S}_{gg}(E_M,\mu_0)
\Big|\lambda_t^{(q)}\C_{1g}\int d u  \phi_M(u)
\frac{\alpha_s}{4\pi} \Big(\frac{1}{u}+\frac{1}{\bar u}\Big)\Big|^2.
\label{two-decay}  
\eeq 
Note that $f_{gg}$ and $\mathcal{S}_{gg}$ are real in contrast to $f_g$ and $\mathcal{S}_g$.

To recapitulate, the prediction for the $B\to \eta(\eta')X$ decay
widths at LO in $1/m_b$ is given by 
\begin{equation}\label{Gammaeta}
\frac{d\Gamma}{dE_M}= \Big(\frac{d\Gamma}{dE_M}\Big)_{\rm
  Eq.\eqref{decay}}+\Big(\frac{d\Gamma}{dE_M}\Big)_{g}+
\Big(\frac{d\Gamma}{dE_M}\Big)_{gg},   
\end{equation} 
with the first term interpreted according to the replacement rules given
explicitly in Eqs.~\eqref{heta}-\eqref{peta'c}, while the last two
terms are given at leading order in $\alpha_s(\sqrt{\Lambda m_b})$ in
Eqs.~\eqref{one-decay} and \eqref{two-decay} and to all orders in
Appendix \ref{app:Gderiv}.

\section{Phenomenology}\label{pheno}
We are now ready to use the expressions for CP averaged branching
ratios and direct CP asymmetries derived in the previous two sections
for quantitative analysis. We split the discussion into two parts,
first focusing on the decays to nonisosinglet pseudoscalar and to
vector final states and then moving to the predictions for the $B\to
\eta X, \eta'X$ decays.  

While the first measurements of $B\to MX$ decays have become available,
one still lacks enough experimental information to determine
nonperturbative charming penguin parameters from data (or to show
decisively that 
they are small and compatible with zero). 
Therefore we collect in Tables \ref{table:DeltaS=1} and
\ref{table:DeltaS=0} only purely perturbative 
predictions for $B\to XM$ decay rates and  direct CP asymmetries
using Eqs.~\eqref{ratioG} 
and \eqref{ACP},  setting the nonperturbative charming penguins
parameters $p_{cc}^M$ and $\cP_{cc}^M$ to zero. Comparison with data
then gives an insight about the importance of nonperturbative charming
penguin contributions and/or  on the size of subleading terms as
detailed below Eqs.~\eqref{K+norm} and \eqref{K0norm}. 
 To reduce the hadronic uncertainties, the predictions
for $B\to XM$ branching ratios are normalized to $d\Gamma(B\to
X\gamma)/dE_\gamma$. The predicted ratio of partial decay widths,
Eq.~\eqref{ratioG}, depends on the light meson energy $E_M$. In the 
endpoint region, however, the dependence on  
$x_M=2E_M/m_B=1+{(m_M^2-p_X^2)}/{m_B^2} $ is a subleading
effect.\footnote{For instance, the same $p_X^2$ cut corresponds to
 higher $E_M$ cut for heavier mesons. For $p_X^2<(2{\rm\ GeV})^2$ one has
$E_\pi>2.26$ GeV for $B\to \pi X$, while $E_\phi>2.36$ GeV for $B\to
\phi X$ (to be compared with 
$m_{B^0}/2=2.64$ GeV). Thus $m_B/2-E_M\sim \Lambda$ with $1-x_M\sim
O(\Lambda/m_B)$.}  
We neglect this dependence and set $x_M=1$ in Tables
\ref{table:DeltaS=1} and \ref{table:DeltaS=0}.

For the coefficients in the Gegenbauer polynomial expansion of the LCDAs
\beq 
\phi_M(u, \mu)=6 u\bar u\left[1+\sum_{n=1}^\infty a_n^M(\mu)
C_n^{3/2}(2u-1)\right],\label{Gegenbauer}
\eeq
we take the same values as  Ref.~\cite{Chay:2006ve}, except for the first
coefficient in the Gegenbauer expansion of  $\phi_K(x)$ for which we
use the recent lattice QCD determination
$a_1^K(\mu=2.0\rm{\ GeV})=0.055\pm0.005$
\cite{Boyle:2006pw}. Explicitly, the remaining coefficients are
(at $\mu=2$ GeV):  $a_2^K=0.23\pm0.23$ \cite{Ball:2006wn},
$a_1^{K^*}=0.08\pm0.13$, \cite{Ball:2004rg}, 
$a_2^\pi=0.09\pm0.15$ \cite{Ball:2004ye}, $a_2^{K^*}=0.07\pm0.08$,
$a_2^\rho=0.14\pm0.15$, $a_2^\phi=0.\pm0.15$ \cite{Ball:2004rg}, and
for lack of better information $a_2^{\eta_q}=a_2^{\eta_s}=a_2^\pi$
and $a_2^\omega=0.\pm0.2$, with the higher coefficients in the
expansion set to zero. 

\begin{table}
\begin{tabular}{lccc}\hline\hline
$MX$ & ${\rm Br}(B^-\to MX )/{\rm Br}(B\to X_s\gamma$)&  $A_{CP}$
\\\hline\hline 
$K^{-} X^0_{u\bar u} $   & $0.17\pm0.09\pm0.06$ &  $0.30\pm0.16\pm0.01$ \\ 
$\overline K^{0} X_{d\bar u}^- $    & $0.20\pm0.11\pm0.06 $&  $(9.7\pm
4.8\pm0.6)\times 10^{-3}$ \\ 
$ \pi^{0} X_{s \bar u}^- $   & $(1.0\pm 0.6\pm0.2)\times 10^{-2}$ &  $-$ \\ 
\hline 
$K^{*-} X^0_{u \bar u} $  & $0.28\pm0.16\pm0.06$ &  $0.32\pm0.16\pm0.02$ \\  
$\overline K^{0*} X_{d\bar u}^- $    & $0.34\pm0.19\pm0.07 $&
$(8.4\pm 4.6\pm1.9)\times 10^{-3}$ \\  
$\phi X_{s\bar u}^- $  & $0.22\pm0.13\pm0.03$ &  $(8.9\pm
5.0\pm1.6)\times 10^{-3}$ \\  
$\omega X_{s\bar u}^- $  & $(2.8\pm 3.3\pm0.7)\times 10^{-3}$ &
$0.49\pm0.24\pm0.33$ \\ 
$ \rho^{0} X_{s \bar u}^- $   & $(2.4\pm 1.4\pm0.5)\times 10^{-2}$ &  $-$ \\ 
\hline  \hline    
\end{tabular}
\caption{\baselineskip 3.0ex 
Predictions for decay rates and direct CP asymmetries for charged $B^-\to MX$
 $\Delta S=1$ semi-inclusive hadronic decays, which are the same as
 for corresponding $\overline B^0\to MX$, $\overline B_s^0\to MX$
 given in Table \ref{table:TMs}. The first errors   
are an estimate of the $1/m_b$ corrections, while the second errors
are due to errors on the Gegenbauer coefficients in the expansion of
the LCDA.} 
\label{table:DeltaS=1}
\end{table}

Direct CP asymmetries, Eq.~\eqref{ACP}, are nonzero only in the
presence of nonzero strong phases. These can be generated
nonperturbatively or by integrating 
out on-shell light quarks in a loop when matching full QCD to $\SI$ at
NLO in $\alpha_s$. As in Ref.~\cite{Chay:2006ve} we therefore use the
NLO matching expressions for the 
Wilson coefficients $\mathcal{C}_i^p$ at $\mu=m_b$, to have the
leading contribution to the CP asymmetries, while performing the 
evolution to the hard-collinear scale $\mu_0\sim \sqrt{\Lambda m_b}$
at NLL. Note that this running cancels to a large extent in
the ratios of the decay rates (only the running of 
$a_n^M(\mu), n\geq 1$ remains), giving in effect
the Wilson coefficients with NLO accuracy at the hard-collinear scale
$\mu_0$ \cite{Chay:2006ve}. We choose $\mu_0=2$ GeV for the
perturbative predictions in Tables  \ref{table:DeltaS=1} and  
\ref{table:DeltaS=0}.

The two errors quoted in Tables \ref{table:DeltaS=1} and
\ref{table:DeltaS=0} are an estimate of subleading corrections 
and due to the errors on the Gegenbauer polynomial coefficients in
the LCDA expansion \eqref{Gegenbauer}. Since the predictions are made to 
NLO in $\alpha_s(m_b)$ but
only to LO in $1/m_b$, the largest corrections are expected to arise
from the $1/m_b$ terms. These are estimated by independently varying
the magnitudes of the leading terms proportional to
$\lambda_{u,c,t}^{(q)}$ by 20\% $\sim O(\Lambda/m_b)$ and the
strong phase by $5^\circ$. This latter variation estimates the error
on the strong phase arising from the uncalculated $\alpha_s(m_b)/m_b$
or $\alpha_s^2(m_b)$ terms. A $100\%$ error 
is assigned to predictions for branching ratios in
color-suppressed tree and  QCD penguin-dominated $\Delta S=0$ decays 
where the $1/m_b$  corrections are sizable compared to the leading
results due to the hierarchy of Wilson coefficients. 
No prediction on CP asymmetries is given for these modes or for the
QCD penguin-dominated $\Delta S=1$ decays.
\begin{table}
\begin{tabular}{lccc}\hline\hline
$MX$ & ${\rm Br}(B^-\to MX )/{\rm Br}(B\to X_s\gamma$)&  $A_{CP}$
\\\hline\hline 
$\pi^{-} X^0_{u\bar u} $   & $0.67\pm0.37\pm0.14$ &  $-0.04\pm0.02\pm0.01$ \\ 
$\pi^0 X_{d\bar u}^- $    & $(4.1\pm 2.1\pm2.6)\times 10^{-3}$&
$0.64\pm0.10\pm0.10$ \\ 
$ K^{0} X_{s \bar u}^- $   & $(1.0\pm 0.5\pm0.3)\times 10^{-2}$ &
$-0.15\pm0.11\pm0.01$ \\  
\hline 
$\rho^- X^0_{u \bar u} $  & $1.76\pm0.97\pm0.38$ &  $-0.04\pm0.02\pm0.01$ \\  
$\rho^0 X_{d\bar u}^- $    & $(1.3\pm 0.6\pm0.7)\times 10^{-2}$&
$0.63\pm0.10\pm0.10$ \\  
$ K^{*0} X_{s \bar u}^- $   & $(1.4\pm 0.8\pm0.5)\times 10^{-2}$ &
$-0.17\pm0.11\pm0.03$ \\ 
$\phi X_{d\bar u}^- $  & $(2.0\pm 1.1\pm0.1)\times 10^{-4}$ &  $-$ \\ 
$\omega X_{d\bar u}^- $  & $(3.8\pm 1.8\pm1.1)\times 10^{-3}$ &
$-0.72\pm0.13\pm0.20$ \\ 
\hline  \hline    
\end{tabular}
\caption{\baselineskip 3.0ex 
Predictions for decay rates and direct CP asymmetries for charged $B^-\to MX$
 $\Delta S=0$ semi-inclusive hadronic decays, which are the same as
 for corresponding $\overline B^0\to MX$, $\overline B_s^0\to MX$
 given in Table \ref{table:TMd}. The first errors   
are an estimate of the $1/m_b$ corrections, while the second errors
are due to errors on the Gegenbauer coefficients in the expansion of
the LCDA.} 
\label{table:DeltaS=0}
\end{table}

Next we confront the perturbative predictions with experimental
data. Normalizing the BaBar results on  semi-inclusive $B\to KX$
branching 
ratios, Eq.~\eqref{KBr}, to ${\rm Br}(B\to
X_s\gamma)=(172\pm21)\times 10^{-6}$  with the same photon energy cut
$E_\gamma>2.34 {\rm\ GeV}$ that was used for the kaon momentum
\cite{Buchmuller:2005zv}, one has  
\begin{align}
\frac{\Gamma(B^-/\overline B^0\to K^- X)}{\Gamma(B\to
  X_s\gamma)}&=1.13\pm0.30,\label{K+norm}\\ 
\frac{\Gamma(B^-/\overline B^0\to \overline K^0 X)}{\Gamma(B\to
  X_s\gamma)}&=0.89\pm0.42.\label{K0norm} 
\end{align}
The central values of the measurements are substantially higher than
the perturbative predictions for $B^-\to K^{-} X^0_{u\bar u} $   and 
$B^-\to \overline K^{0} X_{d\bar u}^- $ modes,\footnote{The
  measurements are an average over charge and neutral $B$ decays to 
$K^-X$ (or $\overline K^0X$) final state, but these are the same to
the order we are working, see Table \ref{table:TMs}. Decays to
$\overline K^0X$ final state include also an incoherent sum with
$\Delta S=0$ decays into $ K^{0} X_{s \bar u}^- $, which are, however,
CKM suppressed and thus small, see Table \ref{table:DeltaS=0}.} given
in Table  \ref{table:DeltaS=1}. They are still consistent within
errors, with the discrepancies respectively at $3\,\sigma$ and
$1.6\,\sigma$ levels for $K^+X$ and $K^0X$ modes, but do indicate that
there might be substantial nonperturbative charming penguin
contributions (or very large $1/m_b$ corrections). 
Using isospin symmetry the charming penguin parameters, $p_{cc}^K$ and
${\cal P}_{cc}^K$, in the two modes are the same. The three real
parameters, $|p_{cc}^K|$,  $\arg(p_{cc}^K)$, and ${\cal P}_{cc}^K$,
can then be determined from the four observables: the two branching
ratios \eqref{K+norm}, \eqref{K0norm} and the corresponding CP
asymmetries once these are measured. This would also leave one
observable as a consistency check.  

Since the CP asymmetries are not measured yet, this procedure is not
possible at present without further approximations.  Quite generally
one expects that roughly $| 
p_{cc}^K|^2 \sim \cP_{cc}^K$. As a starting point we thus take this
relation to be exact and then extract $ |p_{cc}^K|$
as a function of $\arg(p_{cc}^K)$ from the $K^+X$ ($K^0X$) decay
width. This gives for the nonperturbative charming penguin to be about
a factor $4\pm3$ ($4.5\pm2$) larger then the perturbative prediction,
i.e.~the part of $\C_4^c$ containing the $C_1$ Wilson coefficient.

Here the error is a sum of the
experimental error and the error due to the variation of
$\arg(p_{cc}^K)\in [0,2\pi)$. Another way of presenting this result is
through 
the ratio of nonperturbative and perturbative contributions to the
decay width 
\beq
\left|\frac{\lambda_c^{(s)} p_{cc}^K}{h_K^{(s)}}\right|=
\left\{
\begin{matrix}
2.2\pm1.1 &: K^+X\\
2.0\pm1.5 &: K^0X,
\end{matrix}
\right.
\eeq
that should be zero if the charming penguins are purely
perturbative. The error on the ratios mostly reflects the variation
due to a scan over the phase of $p_{cc}^K$. 
These values are very sensitive on the assumed relation between $|
p_{cc}^K|^2 $ and $\cP_{cc}^K$ and should be taken as a rough
guide only. 
Nevertheless, they show that there is experimentally a possible
indication for sizable nonperturbative charming penguin. 
Two ingredients would help to clarify the situation
significantly. First, the inclusion of chirally enhanced $1/m_b$ terms
would show whether part of the discrepancy can be attributed to those
terms \cite{Beneke:1999br,Jain:2007dy}. Second, the 
cut on $p^*(K)$ should be lowered experimentally below the rather high 
value of $2.34$ GeV used at present \cite{Aubert:2006an}, so that one
would be sure that the measurement is in the endpoint region, where
our calculations are applicable, and away from the resonance region.

Further tests are possible once more modes are measured. For instance,
by using SU(3) flavor symmetry the charming penguin parameters for
different modes can be related to only three real parameters, $|p_{cc}|$,
$\arg(p_{cc})$ and ${\cal P}_{cc}$, making 
the framework even more predictive (at the expense of some accuracy
due to SU(3) breaking). Note that in order to realize experimentally
whether there are nonperturbative charming penguins and/or large
$1/m_b$ corrections one does not need any symmetry arguments, just a
comparison between our perturbative predictions and the experiment. To
distinguish between the two sources of corrections, however, the
flavor symmetries would most likely be needed. 

An interesting set of modes that could be used to settle the question
concerning large $1/m_b$ corrections versus nonperturbative charming
penguin are the  
decays where no charming penguins are present. These are the $\Delta
S=0$ decay $B\to \phi X$ and the color suppressed $\Delta S=1$ decays
$B\to \omega X, \pi^0 X$. They are experimentally more challenging
since one would also need to measure the strangeness content of the
inclusive jet. Namely, the related decays, the $\Delta S=1$ decay
$B\to \phi X$ and the $\Delta S=0$ decays $B\to \omega X, \pi^0 X$, do
contain charming penguins and should thus be distinguished
experimentally from the first set of modes that does not contain
charming penguins. Furthermore, the decays that do not receive
charming penguin contributions might have substantial $1/m_b$
corrections \cite{Chay:2006ve}, so one may need to go one higher order
in the $1/m_b$  expansion to have a definite understanding
of experimental results.

We next move to the decays involving $\eta$ and $\eta'$ mesons. The
expressions for the decay widths in the endpoint region have been
derived in Section \ref{sec:glue} with the final result given in
Eq.~\eqref{Gammaeta}. At LO in $1/m_b$, there appear two new shape functions  that
are specific to $B\to \eta^{(')}X$ decays. At lowest order in $\alpha_s$,  ${\cal S}_g(E_M,\mu_0)$ and  
${\cal S}_{gg}(E_M,\mu_0)$ defined in Eqs.~(\ref{tildefg}),  (\ref{fgg}) (or $\mathcal{F}_g^M$ and $\mathcal{F}_{gg}^M$ in Eqs.~(\ref{fg}) and (\ref{fgg:app})  to all orders in $\alpha_s$) describe gluonic contributions coming from one and two
${\cal O}_{1g}$ operator insertions respectively,
cf.~Fig.~\ref{semigluonic}. Little is known about these new
nonperturbative functions, because of the lack of experimental
data. At present only the $B\to \eta'X$ partial decay width with
$E_{\eta'}>2.218 $ GeV cut has been measured
\cite{Browder:1998yb,Aubert:2004eq,Bonvicini:2003aw}. Normalizing to
the $B\to X\gamma$ decay width with the same $E_\gamma$ cut gives\footnote{If
  instead the same $p_{\eta'}$ and $p_\gamma$ cuts are used the decay
  widths ratio is $1.34\pm0.30$.  The difference compared to
  Eq.~\eqref{etaExpratio} 
  reflects the effect of the large $m_{\eta'}$ mass.} 
\beq\label{etaExpratio}
\frac{\Gamma(B\to \eta' X)}{\Gamma(B\to X_s\gamma)}=1.76\pm0.40.
\eeq

In $B\to \eta^{(')} X$ decays there are 8 observables that are
independent at LO 
in $1/m_b$: the $\Delta S=1$ and $\Delta S=0$ $B\to \eta' X, \eta X$
decay widths and direct CP asymmetries. Working at LO in
$\alpha_s(\sqrt{\Lambda m_b})$ and neglecting the $E_M$ dependence of
shape functions, one introduces only two new nonperturbative
parameters specific to these decays, ${\cal S}_g$ 
and ${\cal S}_{gg}$. With more data they could
in principle be determined from experiment in the 
future. Note in particular that the charming penguin parameters
entering the predictions can be fixed using SU(3) flavor symmetry from
semi-inclusive decays into nonisosinglets. 

To have some future guidance on the size of the missing components, we
list in Table \ref{table:eta} the purely perturbative predictions
where we set both the charming penguin as well as the new gluonic
shape function contributions to zero and neglect NLO corrections
from ${\cal O}_{2g}$ insertions. One striking aspect of that
calculation is that the prediction for $\Delta S=1$ $B \to \eta'
X_{s}$ decay falls remarkably short of the large measured
value. Furthermore in this incomplete perturbative prediction the
hierarchy between $\eta X_s$ and $\eta' X_s$ decays is exactly
opposite to the one found in two body $B\to \eta' K, \eta K$
decays. While ${\rm Br}(B\to 
\eta' K)\gg {\rm Br}(B\to \eta K)$, the hierarchy in the incomplete
prediction for the semi-inclusive decays is inverted. In the two body
decays the hierarchy is well explained through the constructive and
destructive interference of $B\to \eta_q K$ and $B\to \eta_s K$
contributions in $B\to \eta' K$ and $B\to \eta K$ amplitudes
respectively  due to $\eta-\eta'$ mixing, if $A(B\to \eta_q K)\simeq
A(B\to \eta_s K)$. This is exactly what is found in the limit of
charming penguin dominance \cite{Williamson:2006hb,Lipkin:1990us}. In
the semi-inclusive decay on the other 
hand there are no charming penguin contributions to $B\to \eta_q X_s$
at LO in $1/m_b$, cf.~Eq.~\eqref{pccDeltaS=1}, 
so that with large charming penguins there is no large hierarchy
between $B \to \eta' X_{s}$ and $B \to \eta X_{s}$ decays. 

To be more quantitative, it is instructive to take three formal
limits: (i) dominating charming penguins, (ii) dominating ${\cal
  O}_{2g}$ contributions, and (iii) the incomplete perturbative
prediction with $\C_{2g}\to 0$. If the amplitudes are dominated by
charming penguins then (in the SU(3) symmetric limit) 
\beq
\frac{{\rm Br}(B^- \to \eta X_{s\bar u}^-)}{{\rm Br}(B^- \to \eta' X_{s\bar
    u}^-)}=\tan^2 \phi=0.67. 
\eeq
Thus, as already argued above, in this limit there is no hierarchy
between the two decays since Lipkin's argument of destructive and
constructive interferences does not work for semi-inclusive decays. On
the other hand, if ${\cal O}_{2g}$ contributions dominate  then
working at LO in $\alpha_s(\sqrt{\Lambda m_b})$ we have  
\beq\label{O2glimit}
\frac{{\rm Br}(B^- \to \eta X_{s\bar u}^-)}{{\rm Br}(B^- \to \eta' X_{s\bar
    u}^-)}=\frac{\big|( \cos \phi f_{\eta_q}\phi_{\eta_q}- \sin \phi
  f_{\eta_s}\phi_{\eta_s})\otimes \big(\frac{1}{u}+\frac{1}{\bar
    u}\big)\big|^2}{\big|( \cos \phi f_{\eta_q}\phi_{\eta_q}+ \sin
  \phi f_{\eta_s}\phi_{\eta_s})\otimes \big(\frac{1}{u}+\frac{1}{\bar
    u}\big)\big|^2}. 
\eeq
Numerically this gives $1.2\times 10^{-4}$ if asymptotic LCDA are used,
and $1.52 \times 10^{-2}$ if SU(3) breaking is estimated by setting
$a_2^{\eta_s}=a_2^{K}$ instead. In the limit of dominant ${\cal
  O}_{2g}$ contributions we thus have a similar large hierarchy
between $\eta X_s$ and $\eta'X_s$ decays as in the two-body decays due
to the destructive interference as apparent from
Eq.~\eqref{O2glimit}. Finally, if the incomplete perturbative calculation
were a valid approximation, then we would have an inverted
hierarchy between $\eta X_s$ and $\eta'X_s$ decays. This is due to a
cancellation that is found between different terms in the $B\to
\eta'X_s$ perturbative prediction. Due to small $B\to \eta'X_s$ decay
width, however, this limit is phenomenologically excluded.  

In order to understand the relative size of charming penguin and the
${\cal O}_{2g}$ contributions it is important to have a measurement of
$B\to \eta X_s$ decays. The relative size compared to the $B\to \eta' X_s$
decay width is clearly different in the two extreme cases when only
one of the two contributions is important. 
Comparing further with the other decays one should be able
to determine all the nonperturbative parameters. As an exercise we
set  $\cP^{\eta_s}=|p_{cc}^{\eta_s}|^2$ and take $p_{cc}^{\eta_s}$ to
be equal to $p_{cc}^K$ obtained from $B\to K^-X$, leading to a
prediction for the normalized decay width
${\rm Br}(B^-\to \eta' X_{s \bar u}^- )/{\rm Br}(B\to
X_s\gamma)=0.43\pm0.25$ with 
the variation mainly due to the unknown strong phase of $p_{cc}^K$
(while for the $\Delta S=0$ decay we find ${\rm Br}(B^-\to \eta' X_{d \bar u}^-
)/{\rm Br}(B\to X_s\gamma)=0.02\pm0.02$). This is still smaller then the
measured value \eqref{etaExpratio}, not surprising given the
approximations made to arrive at it. Whether the difference is
partially explained also by ${\cal 
  O}_{2g}$ contributions should be clarified once more data are
available.

\begin{table}
\begin{tabular}{lccc}\hline\hline
$MX$ & ${\rm Br}(B^-\to MX )/{\rm Br}(B\to X_s\gamma$)&  $A_{CP}$
\\\hline\hline 
$ \eta X_{s \bar u}^- $   & $(5.6\pm 2.9\pm0.6)\times 10^{-2}$ &
$(-7.1\pm 1.8\pm2.8)\times 10^{-2}$ \\  
$ \eta' X_{s \bar u}^- $   & $(1.0\pm 2.0\pm0.3)\times 10^{-2}$ &
$0.19\pm0.19\pm0.08$ \\  
\hline 
$ \eta X_{d \bar u}^- $   & $(6.2\pm 3.2\pm1.3)\times 10^{-2}$ &
$-0.38\pm0.10\pm0.10$ \\  
$ \eta' X_{d \bar u}^- $   & $(2.4\pm 1.2\pm0.7)\times 10^{-2}$ &
$-0.46\pm0.12\pm0.10$ \\  
\hline  \hline    
\end{tabular}
\caption{\baselineskip 3.0ex 
Predictions for decay rates and direct CP asymmetries for charged
$B^-\to \eta X^-,\eta' X^-$ decays  $\Delta S=1(0)$ semi-inclusive
hadronic decays  given above (below) the horizontal line. The predictions
equal also the corresponding $\overline B^0\to \eta X, \eta' X$,
$\overline B_s^0\to \eta X, \eta' X$ decay as given in Table
\ref{table:TMd}. The first errors   
are an estimate of the $1/m_b$ corrections, while the second errors
are due to errors on the Gegenbauer coefficients in the expansion of
the LCDA. 
} 
\label{table:eta}
\end{table}

\section{Conclusions}\label{sec:conclusions}

In the framework of SCET we considered semi-inclusive, hadronic
decays $B\to X M$ in the endpoint region, where the light meson $M$
and the inclusive jet $X$ with $p_X^2\sim \Lambda m_b$ are emitted
back-to-back. This is an extension  of the analysis done in 
Ref.~\cite{Chay:2006ve} where we limited consideration to decays
in which the spectator quark does not enter into the meson $M$. 
The contributions in which the spectator quark enters the meson $M$
are power suppressed by $1/m_b^2$ in SCET. In this work we thus extend
the SCET predictions at LO in $1/m_b$ to all decays 
where the spectator can enter either the jet $X$ or the meson $M$.  
In SCET the four-quark operators factorize, which allows for a
systematic theoretical treatment. After matching the full QCD
effective weak Hamiltonian 
onto $\SI$, the weak interaction four-quark operators factor into the
heavy-to-light current and the $\n$-collinear current. The forward  
scattering amplitude of the heavy-to-light currents leads to a
convolution $\mathcal{S}$ of the jet function with the $B$-meson shape
function, while the matrix element of $\n$-collinear currents gives
the LCDA for the meson $M$. The product of the two then gives the
factorized form for the decay 
rates. The two nonperturbative functions, the convolution
$\mathcal{S}$ and the LCDA, are the only nonperturbative inputs in the
predictions for $B\to X M$ decay rates at leading order in
$1/m_b$. Furthermore, the same convolution $\mathcal{S}$  appears in
$B\to X_s \g$ decay and drops out in the ratio of $B\to X M$ to the
$B\to X_s\g$ rate and in the predictions for direct CP asymmetries.   
Further work on higher order corrections would be useful in reducing the
theoretical uncertainty.

Nonperturbative charming penguin contributions can be included by 
the addition of one real and one complex parameter in the SU(3)
symmetry limit.  These parameters, 
which are zero if the charming penguins are purely perturbative, can
in principle be extracted from data.  Thus by investigating decays
without charming penguins, we can test whether the formalism is
working.  Then by looking at modes where the charming penguin 
can contribute, we can potentially see whether or not the charming
penguin gives a large contribution to the decays.  

Decays where the light meson is an isosinglet $\eta$ or $\eta'$ are special 
in that they receive additional contributions from gluonic operators.  We
consider these decays in detail, and show that the decay rate still factorizes,
but there are two new shape functions which enter into the predicted rate.

Using the available data, we performed an analysis of the semi-inclusive
hadronic decays.  While to date the data  is limited, 
our preliminary analysis seems to indicate either large higher order corrections
or a large contribution from non-perturbative charming penguins.  
With more data, it may be possible to distinguish the two
possibilities as well as to extract the size of the nonperturbative
charming penguin contributions.

\section*{Acknowledgments}
We thank F.~Blanc, I.~Rothstein, and J.~Smith for discussions.
J.~C.~is supported by Grant No. R01-2006-000-10912-0 from the Basic
Research Program of the Korea Science and Engineering Foundation.
C.~K.~is supported in part
by the National Science Foundation under Grant No.~PHY-0244599. 
A.~K.~L.~is supported in part by the National Science Foundation 
under Grant No.~PHY-0546143 and in part  by the Research Corporation.  
The work of J.Z. is supported in part by the European Commission RTN
network, Contract No.~MRTN-CT-2006-035482  
(FLAVIAnet) and by the Slovenian Research Agency. 
J.Z. would like to thank Carnegie Mellon University High Energy
group for hospitality while part of this work was completed.

\appendix
\section{Tree-level matching}\label{app:tlm}
The matching of the effective weak Hamiltonian in full QCD
\beq
H_{W} = \frac{G_F}{\sqrt{2}} 
\Biggl[ \sum_{p=u,c}  \lambda_p^{(q)} \Bigl(C_1 O_1^p + C_2 O_2^p
\Bigr) - \lambda_t^{(q)} \Bigl( \sum_{i=3}^{10} C_i O_i + C_g O_g +
C_{\gamma} O_{\gamma}\Bigr) \Biggr],
\eeq
 onto an $\SI$ one was calculated at NLO in $\alpha_s(m_b)$  
first in  Refs.~\cite{Beneke:1999br}, and then in
Ref.~\cite{Chay:2003ju}, giving the $\SI$ effective weak Hamiltonian
in  Eq.~\eqref{hscet}. Here we list for reader's convenience the tree
level result of the matching  
\begin{eqnarray}
\C_{1,2}^p(v) &=& \delta_{up}\Bigl[ C_{1,2}+\frac{C_{2,1}}{N}
\Bigr]+\frac{3}{2}\Bigl[C_{10,9}+\frac{C_{9,10}}{N}\Bigr],  \nonumber
\\ 
\C_3^p (v) &=& \frac{3}{2} \Bigl[ C_7 +\frac{C_8}{N} \Bigr], \nonumber
\\ 
\C_{4,5}^p(v) &=&  C_{4,3}+\frac{C_{3,4}}{N}
-\frac{1}{2}\Bigl[C_{10,9}+\frac{C_{9,10}}{N} 
\Bigr], \nonumber \\
\C_{6}^p(v) &=& C_{5}+\frac{C_{6}}{N} -\frac{1}{2} \Bigl[
C_{7}+\frac{C_{8}}{N}\Bigr], 
\end{eqnarray}
while the NLO results in our notation can be found in Appendix A of
Ref.~\cite{Chay:2006ve}.

\section{Derivation of gluonic contributions}\label{app:Gderiv}
In this appendix we provide details on the derivation of ${\cal G}(0)$
in Eq.~\eqref{tildeJ-maintext} and extend  
the results for $({d\Gamma}/{dE_M})_{g}$ and $({d\Gamma}/{dE_M})_{gg}$
in Eqs.~\eqref{one-decay} and \eqref{two-decay} 
to all orders in $\alpha_s(\sqrt{\Lambda m_b})$.
We start with the $T$-products  
\beq\label{O1gtprod:app}
{\cal G}_{\xi\xi (cg)}=\langle M X|i \int d^4 x T\big\{ {\cal O}_{1g}(0),
{\cal L}_{\xi\xi(cg)}^{(1)} (x)\big\}|B\rangle, 
\eeq
that were already defined in Eq.~\eqref{O1gtprod-maintext} and are
also shown in Fig.~\ref{glueto}. The subleading SCET Lagrangians
appearing in Eq.~(\ref{O1gtprod:app}) are  
\beq
\begin{split}
{\cal L}_{\xi\xi}^{(1)}=&\bar q_{\bar n}'\big(Y_{\bar n}^\dagger i
\sla D_{\rm us}^\perp Y_{\bar n}\big) \frac{1}{n\cdot P} \big( W_{\bar
  n}^\dagger i \sla D_{\bar n}^\perp W_{\bar n}\big) \frac{\sla n}{2}
q_{\bar n}' +\bar q_{\bar n}' \big( W_{\bar n}^\dagger i \sla D_{\bar n}^\perp
W_{\bar n}\big) \big(Y_{\bar n}^\dagger i \sla D_{\rm us}^\perp Y_{\bar n}
\big)\frac{1}{n\cdot P}  \frac{\sla n}{2} q_{\bar n}', 
\end{split} 
\eeq
where the sum over light quark flavors $q'$ is understood, and
\cite{Bauer:2003mg} 
\beq\label{Lcg}
{\cal L}_{cg}^{(1)}=\frac{2}{g^2} \Tr \big\{ \big[i D_0^\mu, i
D_c^{\perp \nu}\big]\big[i D_{0\mu},W_{\bar n}i D_{{\rm us} \nu}^\perp
W_{\bar n}^\dagger\big]\big\}, 
\eeq
with $i D_0^\mu =i {\cal D}^\mu +g A_{\bar n}^\mu$ and $i{\cal
  D}^\mu=\frac{\bar n^\mu}{2} P+P_\perp ^\mu+\frac{n^\mu}{2} i \bar n
\cdot D_{\rm us}$. 

For the calculation of ${\cal G}_{\xi\xi}$ it is useful to rewrite 
\beq
\begin{split}\label{Lxixi}
{\cal L}_{\xi\xi}^{(1)}={\cal L}_{\xi\xi,a}^{(1)}+{\cal
  L}_{\xi\xi,b}^{(1)}=&\bar q_{\bar n}' g \sla\! {\cal
  A}_{\rm us}^{\perp} \frac{1}{n\cdot P} i g \sla {\cal B}_{\bar n}^\perp
\frac{\sla n}{2} q_{\bar n}' +\bar q_{\bar n}' i g \sla {\cal B}_{\bar
  n}^\perp g \sla\! {\cal A}_{\rm us}^{\perp} \frac{1}{n\cdot P}
  \frac{\sla n}{2}  
q_{\bar n}'+\dots, 
\end{split}
\eeq
with ${\cal A}_{\rm us}^{\perp\mu}$ defined in Eq.~\eqref{A:def}, while
the ellipses denote additional terms containing $P_\perp$ that do 
not contribute in our case. ${\cal G}_{\xi\xi}$ is then also split
accordingly into 
\beq
{\cal G}_{\xi\xi, a(b)}=\langle M X|i \int d^4 x T\big\{ {\cal
  O}_{1g}(0), {\cal L}_{\xi\xi, a(b)}^{(1)} (x)\big\}|B\rangle. 
\eeq
In the SCET$_{\rm I}$ to SCET$_{\rm II}$ matching (where $p^2\sim
 \Lambda m_b$ intermediate degrees of freedom are integrated out) we
focus on the $\n$ fields in ${\cal G}_{\xi\xi,a}$. The matching
 leads to two jet functions once hard-collinear modes are integrated
 out, and we obtain
\beq
\begin{split}\label{Tprod}
T\{[&{\cal B}_{\bar n }^{\perp\mu}]^{cd}(0), [(\bar q_{\bar
  n})^{a}(\gamma_\perp^\alpha \sla {\cal B}_{\bar n}^{\perp} \sla n
q_{\bar n})^b]_u(x)\}=i\delta(x_-)\delta^2(x_\perp) \frac{1}{ m_b} \\ 
&\times\int \frac{d k_-}{2 \pi} e^{-i k_- x_+/2}\big[
(T^A)^{ba}(T^A)^{cd} J_1(u,k_-)+\delta^{ab}\delta^{cd}
J_1'(u,k_-)\big] [\bar q_{\bar n}\gamma_\perp^\alpha \gamma_\perp^\mu
\sla n q_{\bar n}]_u+\cdots, 
\end{split}
\eeq
with $k_-=\n\cdot k$, $x_+=n \cdot x$. The ellipses are
terms which do not contribute to  
$\eta^{(')}$ states. Tree-level matching gives for
the jet functions $J_1(u,k_-)=1/(N k_-)$ and
$J_1'(u,k_-)=0$.  The $J_1'$ term does not contribute
to ${\cal G}_{\xi\xi,a}$ since it leads to $\Tr (g  {\cal
  A}_{\rm us}^{\perp\mu})=0$. The remaining piece can be rearranged using
color identities into 
\beq
\begin{split}
{\cal G}_{\xi\xi,a}&=\frac{\alpha_s}{4\pi} \int \frac{d k_- dx_+}{4\pi} e^{-i
  k_- x_+/2}\int du \frac{J_1(u, k_-)}{u}\\ 
&\times \langle M X|[\bar q_n \sla \bar n \gamma^\perp_\mu P_R
Y_n^\dagger Y_{\bar n} g {\cal A}_{\rm us\,\alpha}^{\perp}(x)Y_{\bar
  n}^\dagger b_v] [\bar q_{\bar n} \gamma_\perp^\alpha
\gamma_\perp^\mu \sla n q_{\bar n}]_u |B\rangle. \label{Gxixiinterm}
\end{split}
\eeq
The two terms in the square brackets are factorized in the sense that
there are no soft gluon exchanges between the two terms -- all 
the soft fields are in the first bracket. The communication between the
two is only through the $k_-$ and $u$ convolutions with the jet
function $J_1(u, k_-)$. 

Using the definition of the LCDA
\beq
\langle M| (q_{\bar n})^a_i [(\bar q_{\bar
  n})^b_j]_u|0\rangle=-\frac{i}{2}E_M f_M \phi_M(u)
\frac{\delta^{ab}}{N}\Big(\frac{\sla \bar n}{2}\gamma_5\Big)_{ij}, 
\eeq
to evaluate the matrix element from the second square bracket in
\eqref{Gxixiinterm} we then have 
\beq
\begin{split}
{\cal G}_{\xi\xi,a}=-\frac{i \alpha_s m_b}{4\pi}\int du  f_M \phi_M(u) 
  &\int \frac{d   k_- dx_+}{4\pi} e^{-i k_- x_+/2} \frac{1}{u} J_1(u,
  k_-)\\  
&\times \langle X|\bar q_n    Y_n^\dagger Y_{\bar n}(0) \sla \bar n g
  \sla\! 
{\cal A}_{\rm us}^{\perp}(x_+)P_R Y_{\bar n}^\dagger b_v(0) |B\rangle. 
\end{split}
\eeq
In simplifying the Dirac structure the identity 
\beq\label{epsident}
\epsilon_\perp^{\alpha \mu}[\sla n, \sla \bar n]\gamma_{\perp \mu}
P_R=4 i \gamma_\perp ^\alpha P_R
\eeq
was used, $\epsilon^{0123} = +1$, and $\epsilon_\perp^{\alpha\mu} =
\epsilon^{\alpha\mu\lambda\sigma}\bar n_\lambda n_\sigma/2$.  For the
derivation of ${\cal G}_{\xi\xi,b}$ we notice 
 that ${\cal L}_{\xi\xi,b}^{(1)}$ is a hermitian conjugate of ${\cal
   L}_{\xi\xi,a}^{(1)}$. Using the hermitian conjugate of
 \eqref{Tprod} we finally have 
\beq
\begin{split}\label{Gxixifinal}
{\cal G}_{\xi\xi}=-\frac{i \alpha_s m_b}{4\pi}\int du f_M \phi_M(u)
  &\int \frac{d 
  k_- dx_+}{4\pi} e^{-i k_- x_+/2} \Big(\frac{1}{u} J_1(u,
k_-)-\frac{1}{\bar u} J_1(\bar u,-k_-)^*\Big)\\ 
&\times\langle X|\bar q_n    Y_n^\dagger Y_{\bar n}(0) \sla \bar n g
  \sla {\cal A}_{\rm us}^{\perp}(x_+)P_R Y_{\bar n}^\dagger b_v(0)
  |B\rangle. 
\end{split}
\eeq

Moving now to the calculation of ${\cal G}_{cg}$, we first rewrite
${\cal L}_{cg}^{(1)}$ in a more useful form 
\beq
{\cal L}_{cg}^{(1)}=\frac{2}{g^2}\Tr\{[ig {\cal B}_{\bar n \perp}^\mu,
ig {\cal B}_{\bar n \perp}^\nu][ig {\cal B}_{\bar n \mu}^\perp, g
{\cal A}_{{\rm us}\nu}^{\perp}]\} 
+\cdots,
\eeq
where the ellipses denote terms that do not contribute to 
${\cal G}_{cg}$. The matching from SCET$_{\rm I}$ to 
SCET$_{\rm II}$ gives 
\beq
\begin{split}
T\{ i g ({\cal B}_{\bar n \perp}^\mu)^{cd}(0), {\cal
  L}_{cg}^{(1)}(x)\} =&\ i \epsilon_\perp^{\mu\nu} g ({\cal A}_{{\rm us}
  \nu}^\perp)^{cd}(x) \delta(x_-)\delta^2(x_\perp) \int \frac{d k_-}{2
  \pi} e^{-i k_-x_+/2}\\ 
&\times\frac{1}{m_b} \int du J_g(u,k_-)
  \epsilon_\perp^{\mu'\nu'}\Tr\big[i g {\cal 
  B}_{\bar n \mu'}^\perp i g {\cal B}_{\bar n \nu'}^\perp\big]_u+\cdots, 
\end{split}
\eeq
where again the ellipses denote terms that do not contribute for
$\eta^{(')}$ final states either because the collinear gluons are not
in color singlet combination or they have incorrect parity. 
Using the definition of gluonic LCDA Eq.~\eqref{gluonicwave} and the
identity \eqref{epsident},  we then have  
\beq
\begin{split}\label{Gcgfinal}
{\cal G}_{cg}=-i \sqrt{C_F} \frac{\alpha_s m_b}{4\pi} \int du  f_P^1
\bar \Phi_P^g(u) &\int \frac{d k_- dx_+}{4\pi} e^{-i k_- x_+/2}
J_g(u,k_-)\\ 
&\times\langle X|\bar q_n    Y_n^\dagger Y_{\bar n}(0) \sla \bar n g
\sla {\cal A}_{\rm us}^{\perp}(x_+)P_R Y_{\bar n}^\dagger b_v(0)
|B\rangle.  
\end{split}
\eeq
 At tree level we have $J_g(u,k_-)=J_g(k_-)$, independent of
 $u$. Since $\bar \Phi_P^g(u)$ is 
antisymmetric, $\bar \Phi_P^g(u)=-\bar \Phi_P^g(\bar u)$, the matrix
element ${\cal G}_{cg}$ vanishes at this order.  The sum of the two
contributions, ${\cal G}_{\xi\xi}$ in Eq.~\eqref{Gxixifinal}  and
${\cal G}_{cg}$ in Eq.~\eqref{Gcgfinal} then gives  
the result for ${\cal G}(0)$ quoted in Eq.~\eqref{tildeJ-maintext}.

We next extend 
the results for $({d\Gamma}/{dE_M})_{g}$ and $({d\Gamma}/{dE_M})_{gg}$
in Eqs.~\eqref{one-decay} and \eqref{two-decay} 
to all orders in $\alpha_s(\sqrt{\Lambda m_b})$. To do so, we redefine
the heavy-to-light current $\tilde {\cal J}$ in  
Eq.~\eqref{tildeJ-integrated} to contain also the integration over
hard momenta fractions $u$ [since  
in general one may not be able to factor this dependence from the
dependence on soft $k_-$ momenta in the jet functions $J_{1,g}(u,
k_-)$], 
\beq\label{JM}
{\cal J}^M(0)=\int du\int \frac{d k_-
  dx_+}{4\pi} e^{-i k_- x_+/2} F^M(k_-,u) \tilde J_H(0,x_+).
\eeq
The heavy current $\tilde J_H$ is given in Eq.~\eqref{JH-maintext}, while the
hard-collinear kernel multiplied by LCDA, $F^M(k_-,u)$, is given in  
Eq.~\eqref{F^M}. Unlike the current $\tilde {\cal J}$ in
Eq.~\eqref{tildeJ-integrated}, the current ${\cal J}^M$ in \eqref{JM}
depends 
on the final state meson $M$ through LCDA that are part of the
$F^M(k_-,u)$ function.

The derivation of the $B\to XM$ decay width is now very similar to the
one given in Section \ref{sec:glue}. 
Starting from the $T$ product of heavy currents corresponding to one
${\cal O}_{1g}$ insertion, 
\beq
T_g^M(E_M)=\frac{i}{m_b}\int d^4 z \langle \overline B| T 
J_H^\dagger(z) {\cal J}^M(0) 
|\overline B\rangle,  
\eeq
with $J_H(z)=e^{i(\tilde p-m_b v)\cdot z}(\bar q_n
\sla \bar n P_L Y_n^\dagger b_v)(z)$ and ${\cal J}^M(0)$ given in
\eqref{JM}, we use the factorization of  
 $n$ collinear quark fields from the rest at LO in $1/m_b$  to write 
\begin{eqnarray}\label{fg}
{\rm Disc.} \  T_g^M(E_M) &=& 2\negmedspace \int\negmedspace dl_+
dr_-\negthinspace\int\negmedspace du \Im
\Bigl[\frac{-1}{\pi}J_P(l_++m_b-2E_M +i \epsilon)\Bigr] \nonumber \\
&\times&  f_g(l_+,r_-)   F^M(r_-,u)  \equiv
2 {\cal F}_g^M(E_M,\mu_0 ), 
\end{eqnarray}
where the $n$-collinear jet function $ J_P(\kappa_++i\epsilon)$ was
defined in Eq.~\eqref{jet-func}, while  the   
 shape function $f_g(l_+,r_-)$ was defined in Eq.~\eqref{f_g2}.
Using the optical theorem we now have  for the decay width contribution
from single ${\cal O}_{1g}$ insertion 
\begin{eqnarray}
\Big(\frac{d\Gamma}{dE_M}\Big)_g&=&\frac{G_F^2}{4\pi} m_b^2 x_M^2 f_M
2\Re\left[\lambda_t^{(q)}\C_{1g} {\cal  F}^M_g(E_M,\mu_0)   \left( \phi_M  \otimes
  \lambda_p^{(q)}T_{M,p}^{(q)}\right)^* 
\right].
\end{eqnarray}
This extends Eq.~\eqref{one-decay} to all orders in 
$\alpha_s(\sqrt{\Lambda m_b})$. We reiterate that the ``shape"
function ${\cal F}_g^M(E_M,\mu_0)$ now contains an integral over hard
momenta fractions in the LCDA so that it depends on the meson $M$. 

Defining similarly for the double ${\cal O}_{1g}$ insertion
\beq\label{Tgg}
T_{gg}^M(E_M)=\frac{i}{m_b}\int d^4 z \langle \overline B| T {\cal
  J}^{M\dagger}(z) {\cal J}^M(0)|\overline B\rangle, 
\eeq
we have
\begin{eqnarray} \label{fgg:app}
{\rm Disc.} \  T_{gg}^M(E_M) &=& 2 \int dl_+ dr_-ds_-\Im
\big[-\frac{1}{\pi}J_P(l_++m_b-2E_M +i \epsilon ) \big]\\ 
&\times&\int du F^M(r_-,u) \int dv F^M(s_-,v)^* f_{gg}(l_+,r_-,s_-)
\equiv 2m_b {\cal F}_{gg}^M(E_M,\mu_0), \nonumber
\end{eqnarray}
where the shape function $f_{gg}(l_+,r_-,s_-)$ that depends on three
soft momenta was defined in Eq.~\eqref{f-three}. 
For the double ${\cal O}_{1g}$ insertion contribution to the
decay width we then have 
\beq
\Big(\frac{d\Gamma}{dE_M}\Big)_{gg}=\frac{G_F^2}{2\pi} m_b^2 x_M
    {\cal   F}^M_{gg}(E_M,\mu_0) \big|\lambda_t^{(q)}\C_{1g} \big|^2,  
\eeq 
which extends Eq.~\eqref{two-decay} to  all orders in
$\alpha_s(\sqrt{\Lambda m_b})$, with the ``shape" function ${\cal
  F}_{gg}^M(E_M,\mu_0)$ again 
depending on the meson $M$ through the LCDA.

\end{document}